\documentclass[11pt]{article}
\usepackage{amssymb}
\usepackage{amsfonts}
\usepackage{graphicx}
\usepackage{amsmath}
\usepackage{hyperref}

\makeatletter \@addtoreset{equation}{section} \makeatother

\newtheorem{theorem}{Theorem}[section]
\newtheorem{lemma}{Lemma}[section]

\newtheorem{remark}{Remark}[section]
\newtheorem{definition}{Definition}[section]
\newtheorem{proposition}{Proposition}[section]

\newcommand{\mdet}{\mathrm{det}}

\newcommand{\Tr}{\mathrm{Tr}\,}

\begin{document}
\title{Characteristic polynomials for 1D random band matrices from the localization side}
\author{ Mariya Shcherbina
\thanks{Institute for Low Temperature Physics, Kharkiv, Ukraine, e-mail: shcherbina@ilt.kharkov.ua} \and
 Tatyana Shcherbina
\thanks{School of Mathematics, Institute for Advanced Study, Princeton, USA, e-mail: tshcherbina@ias.edu. Supported by NSF grant DMS-1128155.}}

\date{}
\maketitle

\begin{abstract}
We study the special case of
$n\times n$ 1D Gaussian Hermitian random band matrices,  when the covariance of the elements is determined by 
$J=(-W^2\triangle+1)^{-1}$. Assuming  that  the band width $W\ll \sqrt{n}$, 
we prove that the limit of the normalized second mixed moment of characteristic polynomials (as $W, n\to \infty$) is equal to one, and so it does not
coincides with those for GUE. This complements the result of \cite{TSh:14} and proves the expected crossover for 1D Hermitian random 
band matrices at $W\sim \sqrt{n}$ on the level of characteristic polynomials.

\end{abstract}
\section{Introduction}
As in \cite{TSh:14}, we consider Hermitian $n\times n$ matrices $H_n$
  whose entries $H_{ij}$ are random
complex Gaussian variables with mean zero such that
\begin{equation}\label{ban}
\mathbf{E}\big\{ H_{ij}H_{lk}\big\}=\delta_{ik}\delta_{jl}J_{ij},
\end{equation}
where
\begin{equation}\label{J}
J_{ij}=\left(-W^2\Delta+1\right)^{-1}_{ij},
\end{equation}
and $\Delta$ is the discrete Laplacian on $\mathcal{L}=[1,n]\cap \mathbb{Z}$ with periodic boundary conditions.
It is easy to see that  the variance of matrix elements $J_{ij}$ is exponentially small when $|i-j|\gg W$, and so  $W$ can be considered as the width of the band.

The density of states $\rho$ of the ensemble is given by the well-known Wigner semicircle law (see
\cite{BMP:91, MPK:92} ):
\begin{equation}\label{rho_sc}
\rho(\lambda)=(2\pi)^{-1}\sqrt{4-\lambda^2},\quad \lambda\in[-2,2].
\end{equation}
Random band matrices (RBM) are natural intermediate models
to study eigenvalue statistics
and quantum propagation in disordered systems, since they interpolate between mean-field type Wigner
matrices (Hermitian or real symmetric matrices with i.i.d.  random entries) and random Schr$\ddot{\hbox{o}}$dinger operators, 
which have only a random diagonal potential in addition to the deterministic Laplacian on a box in $\mathbb{Z}^d$. In particular, 
RBM can be used to model the Anderson metal-insulator phase transition. 

Let $\ell$ be the localization length,
which describes the typical length scale of the eigenvectors of random matrices.
The system is called delocalized if $\ell$ is
comparable with the matrix size, and it is called localized otherwise.
Delocalized systems correspond to electric
conductors, and localized systems are insulators.

According to the physical conjecture (see \cite{Ca-Co:90, FM:91}) for 1D RBM the expected order of $\ell$ is
 $W^2$ (for the energy in the bulk of the spectrum), which means that varying $W$
we can see the crossover: for $W\gg \sqrt{n}$ the eigenvectors are expected to be
delocalized, and for $W\ll \sqrt{n}$ they are localized. 

The questions of the localization length   are closely related to the universality conjecture
of the bulk local regime of the random matrix theory. The bulk local regime deals with the behaviour of eigenvalues of $n\times n$
random matrices on the intervals whose length is of the order $O(n^{-1})$.
According to the Wigner -- Dyson universality conjecture, this local behaviour
does not depend on the matrix probability
law (ensemble) and is determined only by the symmetry type of matrices.
In this language the conjecture about the crossover for 1D RBM states that we get the same behaviour
of eigenvalues correlation functions  as for GUE (Hermitian matrices with i.i.d Gaussian entries) for $W\gg \sqrt{n}$ (which corresponds to delocalized states), and
we get another behaviour determined by the Poisson statistics, for $W\ll \sqrt{n}$
(and corresponds to localized states). 

At the present time only some upper and lower
bounds for $\ell$ are proved rigorously. It is known from the paper \cite{S:09} that $\ell\le W^8$.
On the other side, for the general Wigner matrices (i.e., $W=n$) the
bulk universality has been proved in \cite{EYY:10, TV:11}, which gives $\ell \ge W$. 
By the developing the Erd\H{o}s-Yau approach,  there were also obtained some other results, where the localization length is controlled in a rather weak sense, i.e.
the estimates hold for ``most'' eigenfunctions only: $\ell \ge W^{7/6}$ in \cite{EK:11} and $\ell\ge W^{5/4}$ in \cite{Yau:12}.
Gap universality for $W\sim n$ was proved very recently in \cite{BEYY:16}.

Another method, which allows to work with random operators with non-trivial spatial structures, is supersymmetry techniques (SUSY) based
on the representation of the determinant as an integral over the Grassmann variables. 
This method is widely  used in
the physics literature  and is potentially very powerful but the rigorous
control of the integral representations, which can be obtained by this method, is quite difficult.
The rigorous application of SUSY to the Gaussian RBM  which has the special block-band structure (special case of Wegner's orbital model)
was developed in \cite{TSh:14_1}, where the universality of the bulk local regime for $W\sim n$ was proved.
Combining this approach with Green's function comparison strategy it has been proved recently in \cite{EB:15} that  $\ell \ge W^{7/6}$
(in a strong sense) for the block band matrices with rather general element's distribution.
However, in the general case
of RBM the question of bulk universality of local spectral statistics or of the order of the localization length
is still open even for $d=1$.

Instead of eigenvalues correlation functions one can consider more simple objects which are  the
correlation functions of characteristic polynomials:
\begin{equation}\label{F_2k}
F_{2k}(\Lambda)=\mathbf{E}\Big\{\prod\limits_{s=1}^{2k}\mdet(\lambda_s-H_n)\Big\},
\end{equation}
where  $\Lambda=\hbox{diag}\,\{\lambda_1,\ldots,\lambda_{2k}\}$ are real  parameters
that may depend on $n$. We are interested in the asymptotic behaviour of this function for
\begin{equation}\label{lam_j}
\lambda_j=E+\dfrac{\xi_j}{n\rho(E)},\quad E\in (-2,2).
\end{equation}
From the SUSY point of view, correlation functions of characteristic polynomials correspond to the so-called fermion-fermion sector of the supersymmetric full model
describing the usual correlation functions. They are especially convenient for the SUSY approach and were successfully studied by this techniques
for many ensembles (see \cite{Br-Hi:00}, \cite{Br-Hi:01}, \cite{TSh:11}, \cite{TSh:11_1}, etc.).
Although $F_{2k}(\Lambda)$ is not a local object, it is also expected 
to be universal in some sense. Moreover, correlation functions of characteristic polynomials are expected to exhibit a crossover which is similar to that
of local eigenvalue statistics. In particular, for  1D RBM they are expected to have the same local behaviour 
as for GUE for $W\gg \sqrt{n}$, and the different behaviour for $W\ll \sqrt{n}$. The first part of this conjecture was proved in \cite{TSh:14}.
The main result of \cite{TSh:14} is
\begin{theorem}[\textbf{\cite{TSh:14}}]\label{thm:old}
For the 1D RBM of (\ref{ban})  -- (\ref{J}) with $W^2=n^{1+\theta}$, where $0<\theta\le 1$, we have
\begin{equation}\label{lim}
\lim\limits_{n\to\infty}
D_2^{-1}F_{2}\Big(E+\dfrac{\xi}{n\rho(E)},E-\dfrac{\xi}{n\rho(E)}\Big)
=\dfrac{\sin(2\pi\xi)}{2\pi\xi},
\end{equation}
i.e. coincides with those for GUE. The limit is uniform in $\xi$ varying in any compact set $C\subset\mathbb{R}$. Here
$\rho(x)$ and $F_2$ are defined in (\ref{rho_sc}) and (\ref{F_2k}),
$E\in (-2,2)$, and
\begin{equation}\label{D_2}
D_2=F_2(E,E).
\end{equation}
\end{theorem}
The purpose of the present paper is to study correlation functions of characteristic polynomials for (\ref{ban}) from the localization side $W\ll \sqrt{n}$ and to prove,
that (\ref{lim}) is different  in this case.
The main result is 
\begin{theorem}\label{thm:new}
For the 1D RBM of (\ref{ban}) -- (\ref{J})  with $1 \ll W\le \sqrt {n /C_*\log n}$ for sufficiently big $C_*$, we have
\begin{equation*}
\lim\limits_{n\to\infty}
D_2^{-1}F_{2}\Big(E+\dfrac{\xi}{n\rho(E)}, E-\dfrac{\xi}{n\rho(E)}\Big)
=1,
\end{equation*}
where the limit is uniform in $\xi$ varying in any compact set $C\subset\mathbb{R}$. Here $E\in (-2,2)$, 
and $\rho(x)$, $F_2$, and $D_2$ are defined in (\ref{rho_sc}),  (\ref{F_2k}), and (\ref{D_2}).
\end{theorem}
This theorem complements the previous one and proves the crossover  of the bulk local regime of the random matrix theory on the level of
correlation functions of characteristic polynomials. 
\begin{remark}
Although the result is formulated for $\xi_1=-\xi_2=\xi$ in (\ref{lam_j}), one can prove Theorem \ref{thm:new} for $\xi_1,\xi_2\in C\subset \mathbb{R}$ by the
same arguments with minor revisions. The only difference is a little bit more complicated expressions for $D_2$ and  $K(\xi)$ (see (\ref{Ker}) below).
\end{remark}
To prove Theorem \ref{thm:new}, we apply the transfer matrix approach to the integral representation obtained in \cite{TSh:14} by the supersymmetry techniques
(note that the integral representation does not contain Grassmann integrals, see Proposition \ref{p:int_repr}). 
The main difficulty here is that the transfer operator $K(\xi)$, obtained from an integral representation (see (\ref{Ker}) below), is not self-adjoint; thus perturbation theory is not
easily applied in a rigorous way.  One possible way to work with similar operators was suggested in \cite{D-S:15}, where the much simpler toy-version of $K(\xi)$ (not in the matrix space,  and with one saddle point only) was studied. Here we propose another approach, which does not require the contour rotation.

The paper is organized as follows. In Section $2$ we rewrite $F_2$ as a trace of the $n$-th degree of some transfer operator $K(\xi)$ (see (\ref{Ker}) below) and reduce 
Theorem \ref{thm:new} to the statements on the top eigenvalues of the operator (see (\ref{main_1}), (\ref{main_2})). These statements are proved in Section $4$ (see Theorem
\ref{t:K}). Section $3$ deals with the most important preliminary results needed for Section $4$. The proofs of some technical lemmas are given in Appendix. 

\section{ Representation in the operator form}
As it was proved in \cite{TSh:14}, Lemma 1, we have
\begin{proposition}[\textbf{\cite{TSh:14}}]\label{p:int_repr}
The second correlation function of the characteristic polynomials for 1D Hermitian Gaussian band
matrices, defined in (\ref{F_2k}), can be represented as follows:
\begin{align*}
&F_2\Big(\Lambda_0+\dfrac{\hat{\xi}}{n\rho(E)}\Big)=-(2\pi^2)^{-n}\mdet^{-2} J
\int\exp\Big\{-\frac{W^2}{2}\sum\limits_{j=1}^n\Tr
(X_j-X_{j-1})^2\Big\}\\ \notag
&\times\exp\Big\{-
\frac{1}{2}\sum\limits_{j=1}^n \Tr\Big(X_j+\frac{i\Lambda_0}{2}+
\frac{i\widehat{\xi}}{n\rho(E)}\Big)^2\Big\}\prod\limits_{j=1}^n
\det\big(X_j-i\Lambda_0/2\big)\prod\limits_{j=1}^ndX_j,
\end{align*}
where $\hat\xi=\hbox{diag}\,\{\xi,-\xi\}$, $\Lambda_0=E\,\mathbf{I}$, $X_j\in \hbox{Herm}(2)$ (i.e., $2\times 2$ Hermitian matrices), $X_0=X_n$, and
\begin{equation*}
dX_j=d(X_{j})_{11}d(X_{j})_{22}d\Re (X_{j})_{12}d\Im (X_{j})_{12}.
\end{equation*}
\end{proposition}
Denote 
\begin{equation}\label{H}
\mathcal{H}= L_2[\hbox{Herm}(2)].
\end{equation}
Let $\mathcal{F}:\mathcal{H}\to \mathcal{H}$,  $\mathcal{F}(\xi):\mathcal{H}\to \mathcal{H}$ be the operators of multiplication by
\begin{align}\label{F_cal}
\mathcal{F} (X)&=\exp\Big\{-
\frac{1}{4}\, \Tr\Big(X+\frac{i\Lambda_0}{2}\Big)^2+\frac{1}{2}\,\Tr \log
\big(X-i\Lambda_0/2\big)-F_*\Big\},\\ \notag
\mathcal{F}_\xi (X)&=\mathcal{F} (X)\cdot \exp\Big\{-
\frac{i}{2n\rho(E)}\, \Tr X\hat\xi\Big\}, 
\end{align}
respectively, where $F_*$ will be chosen below (see (\ref{F*})).
Let also $K, K(\xi): \mathcal{H}\to\mathcal{H}$ be the operators with the kernels 
\begin{align} \label{K}
K(X,Y)&=\dfrac{W^4}{2\pi^2}\,\mathcal{F}(X)\,\exp\Big\{-\frac{W^2}{2}\Tr
(X-Y)^2\Big\}\,\mathcal{F}(Y);\\
\label{Ker}
K_{\xi}(X,Y)&=\dfrac{W^4}{2\pi^2}\,\mathcal{F}_\xi(X)\,\exp\Big\{-\frac{W^2}{2}\Tr
(X-Y)^2\Big\}\,\mathcal{F}_\xi(Y).
\end{align}
 Define 
\begin{equation*}
C_n(\xi)=\exp\big\{2nF_*+\xi^2/n\rho(E)^2\big\}
\end{equation*}
Then Proposition \ref{p:int_repr} can be reformulated as
\begin{align}\label{F_rep}
F_2\Big(\Lambda_0+\dfrac{\hat{\xi}}{n\rho(E)}\Big)=-C_n(\xi)\cdot W^{-4n}\mdet^{-2} J\cdot \Tr K^n(\xi),
\end{align}
and thus we are interested in the asymptotic behaviour of $ \Tr K^n(\xi)$.

For arbitrary compact operator $M$ we  denote $\lambda_j(M)$ the $j$th (by its modulo) eigenvalue
of $M$,
so that $|\lambda_0(M)|\ge|\lambda_1(M)|\ge\dots$.

Assume that we have proved that
\begin{align}\label{main_1}
\Big|\frac{\lambda_1(K(\xi))}{\lambda_0(K(\xi))}\Big|\le e^{-C_1/W^2},\quad|\lambda_0(K(\xi))|= 1-C_2/W+O(W^{-2}).
\end{align}
 Then
\begin{align}\label{tr}
&\Tr K^{n}( \xi)=\lambda_0^n(K(\xi))(1+r),
\end{align}
where
\begin{align*}
&|r|=\Big|\sum_{j=1}^\infty\big(\lambda_j(K(\xi))/\lambda_0(K(\xi))\big)^n\Big|\le\Big|\frac{\lambda_1(K(\xi))}{\lambda_0(K(\xi))}\Big|^{n-2}
\sum_{j=0}^\infty|\lambda_j(K(\xi))|^2\\
\le &Ce^{-Cn/ W^2}\int|K_\xi(X,Y)|^2dXdY\le CW^4e^{-CC_*\log n}=o(1),\quad W\to\infty.
\end{align*}
Similarly, 
\begin{align*}
&D_2=-C_n(0)\cdot W^{-4n}\mdet^{-2} J\cdot \lambda_0^n\big(K(0)\big)\cdot(1+o(1))\\
\end{align*}
Thus, the assertion of Theorem \ref{thm:new} follows from (\ref{F_rep}) and (\ref{tr}) combined with the relation
\begin{align}\label{main_2}
|\lambda_0(K(\xi))-\lambda_0\big(K(0)\big)|=o(n^{-1}),\quad n\to\infty.
\end{align}

\section{Preliminary results}
To prove (\ref{main_1}), consider stationary  points of the function $\mathcal{F}$ of (\ref{F_cal}). It is easy to see that they are
\begin{align}\label{st_points_1}
X_+&=a_+\mathbf{I},\quad X_-=a_-\mathbf{I};\\
X_\pm(U)&=a_+\,ULU^*,\quad U\in \mathring{U}(2), \notag
\end{align}
where $ \mathring{U}(2)=U(2)/U(1)\times U(1)$,
\begin{equation}\label{a_pm}
a_+=-a_-= \sqrt{1-E^2/4},\quad L=
\left(
\begin{array}{cc}
1&0\\
0&-1
\end{array}
\right).
\end{equation}
Choose now $F_*$ of (\ref{F_cal}) as 
\begin{align}\label{F*}
F_*=\frac{1}{4}\, \Tr\Big(a_+I+\frac{i\Lambda_0}{2}\Big)^2-\frac{1}{2}\,\Tr \log
\big(a_+I-i\Lambda_0/2\big).
\end{align}
Then the value of $|\mathcal{F}|$ at points (\ref{st_points_1}) is $1$.

Put
\begin{equation*}
X=
\left(
\begin{array}{cc}
a_1&(x_1+iy_1)/\sqrt{2}\\
(x_1-iy_1)/\sqrt{2}&b_1
\end{array}
\right), \quad Y=
\left(
\begin{array}{cc}
a_2&(x_2+iy_2)/\sqrt{2}\\
(x_2-iy_2)/\sqrt{2}&b_2
\end{array}
\right).
\end{equation*}
Rewrite $K(X,Y)$, $K_\xi (X,Y)$  as
\begin{align}
\label{rep_1}
&K_\xi (X,Y)=K(X,Y)+\widetilde K (X,Y),\\
\notag
&K(X,Y)= A(a_1,a_2)\,A(b_1,b_2)\,A_{1}(X,Y),
\end{align}
where the kernels $A$ and $A_1$ have the form
\begin{align}
\label{A}
&A(x,y)=F(x)B(x,y)F(y), \quad B(x,y)=(2\pi)^{-1/2}We^{-W^2(x-y)^2/2};\\ \notag
&F(x)=e^{-f(x)/2},\quad f(x)=(x+iE/2)^2/2-\log (x-iE/2)-F_*;\\ \label{A_1}
&A_1(X,Y)=  F_1(X)B(x_1,x_2)B(y_1,y_2)\,F_1(Y);\\ \notag
&F_1(X)=\exp\Big\{-\dfrac{1}{4}(x_1^2+y_1^2)+\dfrac{1}{2}\log\Big(1-\dfrac{x_1^2+y_1^2}{2(a_1-iE/2)(b_1-iE/2)}\Big)\Big\}, \notag
\end{align}
and the perturbation kernel $\widetilde K$ is
\begin{align*}
\widetilde K(X,Y)=A(a_1,a_2)\,A(b_1,b_2)\,A_{1}(X,Y)\,\Big(e^{-\frac{i}{2n\rho(E)}\big(\xi (a_1-b_1) +\xi (a_2-b_2)\big)}-1\Big).
\end{align*}
Note that
\begin{align}
& f(a_+)=\Re f(a_-)=f'(a_+)= f'(a_-)=0,\quad f(a_{\pm}+x)-f(a_\pm)=c_{\pm}x^2+c_{3\pm}x^3+\dots;\notag\\
&c_\pm=a_+(\sqrt{4-E^2}\pm iE)/2,\quad \Re c_{+}=\Re c_{-}>0;\label{c_pm}\\
&\|B\|\le 1,\quad \| A\|\le 1.\label{norm_A}
\end{align}
Another representation of $K(X,Y)$, $\widetilde{K}(X,Y)$ can be obtained by using polar coordinates. Namely, changing the variables
\[X=U\Lambda U^*,\quad \Lambda=\mathrm{diag}\{a,b\},\quad a>b,\quad U\in \mathring{U}(2),\]
we obtain that $K(\xi)=K+\widetilde{K}$ can be  represented as an integral operator in $L_2[\mathbb{R}^2,p]\times L_2[\mathring{U}(2),dU]$ defined by the kernel
\begin{align}\label{rep_2}
&K_\xi(X,Y)=K(a_1,a_2,b_1,b_2,U_1,U_2)+\widetilde K(a_1,a_2,b_1,b_2,U_1,U_2),
\end{align}
where
\begin{align} \notag
&K(a_1,a_2,b_1,b_2,U_1,U_2)=t^{-1}A(a_1,a_2)A(b_1,b_2)K_*(t,U_1,U_2);\\ \label{K(U)}
&K_*(t,U_1,U_2):=W^2t\cdot e^{tW^2\Tr U_1U_2^*L(U_1U_2^*)^*L/4-tW^2/2};\\
&\widetilde K(a_1,a_2,b_1,b_2,U_1,U_2)=K(a_1,a_2,b_1,b_2,U_1,U_2)\big(e^{(\nu(a_1,b_1,U_1)+\nu(a_2,b_2,U_2))/n}-1\big);\notag\\
&\nu(a,b,U)=-\frac{i\xi\,(a-b)}{4\rho(E)}\,\Tr ULU^*L.
\notag\end{align}
Here and everywhere below 
\begin{align}\label{t}
t=(a_1-b_1)(a_2-b_2), \quad p(a,b)=\dfrac{\pi}{2}(a-b)^2
\end{align}  
and we denote by $dU$ the integration with respect the Haar measure
on the group $\mathring{U}(2)$.
The scalar product and the action of an integral operator in $L_2[\mathbb{R}^2,p]\times L_2[\mathring{U}(2),dU]$ are
\begin{align}\label{(,)_a}
&(f,g)_p=\int f(a,b)\bar g(a,b)p(a,b)dadb,\quad p(a,b)=\dfrac{\pi}{2}(a-b)^2;\\
&(Mf)(a_1,b_1,U_1)=\int M(a_1,a_2,b_1,b_2,U_1,U_2)\,f(a_2,b_2,U_2)\,p(a_2,b_2)da_2db_2dU_2.
\notag\end{align} 
Now let us study the operators $A$ and $K_*$ appearing in (\ref{A}) and  (\ref{K(U)})

\subsection{Analysis of the operator $A$}


\begin{theorem}\label{t:A} 
Operator $A$ of (\ref{A}) has exactly one eigenvalue in  each of  the $CW^{-3/2}$-neighbourhoods
of $\lambda_{0,+}$ and $\lambda_{0,-}$, where
\begin{align}\label{tA.1}
&\lambda_{0,\pm}=\Big(1+\dfrac{2\alpha_\pm}{W}+\dfrac{c_\pm}{W^2}\Big)^{-1/2};\\
\label{alpha}
& \alpha_\pm=\sqrt{\dfrac{c_\pm}{2}}\Big(1+\frac{c_\pm}{2W^2}\Big)^{1/2}.
 \end{align}
Moreover, $|\lambda_2(A)|\le |\lambda_{0,+}|-c_1/W$ with some absolute $c_1>0$.
\end{theorem}
The proof of the theorem is based on  the proposition, which is the standard  linear algebra tool 
 \begin{proposition}\label{p:sp}
Given a compact operator $\mathcal{K}$, assume that there is an orthonormal basis $\{\Psi_l\}_{l\ge 0}$ such that
the resolvent
\[\widehat{\mathcal{G}}_{jk}(z)=(\widehat{\mathcal{K}}-z)^{-1}_{jk},\quad
\widehat{\mathcal{K}}=\{\mathcal{K}_{jk}\}_{j,k=1}^{\infty}\]
is uniformly bounded in   $z\in\Omega\subset \mathbb{C}$, where $\Omega$ is some domain.
Then
the eigenvalues of $\mathcal{K}$ in $\Omega$   coincide with zeros of the function
\begin{align}\label{sp.1}
F(z):=&\mathcal{K}_{00}-z-(\widehat{\mathcal{G}}(z)\kappa,\kappa^{*}), \\
\kappa=&(\mathcal{K}_{10},\mathcal{K}_{20},\dots),\,\kappa^*=(\mathcal{K}^*_{10},\mathcal{K}^*_{20},\dots)
\notag\end{align}
\end{proposition}
The proof of the proposition  follows  from the standard Schur inversion  formula
\begin{align}\label{p_sp.3}
&\mathcal{G}_{ij}(z)=\widehat{\mathcal{G}}_{ij}(z)+(\widehat{\mathcal{G}}\kappa)_i(\widehat{\mathcal{G}}\kappa^*)_j/F(z),\quad i\not=0,j\not=0,\\
&\mathcal{G}_{0j}(z)=-(\widehat{\mathcal{G}}\kappa^*)_j/F(z), \quad \mathcal{G}_{i0}(z)=-(\widehat{\mathcal{G}}\kappa)_i/F(z),\quad
\mathcal{G}_{00}(z)=(F(z))^{-1},
\notag\end{align}
valid for any $z:F(z)\not =0$.

\subsubsection{Proof of  Theorem \ref{t:A}}
To apply the proposition, let us first introduce and study the ``model" operator
\begin{align*}
&A_*^{(c_*)}(x,y)=\mathcal{F}_*(x)B(x,y)\mathcal{F}_*(y),\quad \mathcal{F}_*(x)=e^{-c_*x^2/2}, \quad \Re c_*>0.
\end{align*}
Take
\begin{align*}
 &\alpha=\sqrt{\dfrac{c_*}{2}}\Big(1+\frac{c_*}{2W^2}\Big)^{1/2}=:\alpha_1+i\alpha_2,
\end{align*}
 and consider the system of
 functions
\begin{align}\label{pA.1}
&\psi_0(x)=e^{-\alpha W x^2}\sqrt[4]{\alpha W/\pi},\\
&\psi_k(x)=h_k^{-1/2}e^{-\alpha Wx^2} e^{2\alpha_1W x^2}\Big(\frac{d}{dx}\Big)^ke^{-2\alpha_1W x^2},
=e^{-\alpha Wx^2}p_k(x)\notag\\
&h_k=k!(4\alpha_1W)^{k-1/2}\sqrt{2\pi},\quad k=1,2,\ldots
\notag\end{align}
It is easy to see that $\{p_k\}_{k=0}^\infty$ are polynomials, orthogonal with the weight $e^{-2\alpha_1Wx^2}$
($p_k$ is the $k$th Hermite polynomial of $x\sqrt{2\alpha_1W}$ with a proper normalization).
\begin{lemma}\label{l:A_*}
 $\{\psi_k\}_{k\ge 0}$ is a orthonormal system in $L_2(\mathbb{R})$ and
\begin{align}\label{pA.0}
A_*^{(c_*)}\psi_0=\lambda_0^{(c_*)}\psi_0,\quad \lambda_0^{(c_*)}=\Big(1+\frac{2\alpha}{W}+\frac{c_*}{W^2}\Big)^{-1/2},
\end{align}

The matrix $A_{*jk}^{(c_*)}:=(A_*^{(c_*)}\psi_k,\psi_j)$ is upper triangular, $(A_*^{(c_*)})_{jk}=0$ if $j$, $k$ have
 different evenness, and
\begin{align}\label{pA.3}
&A_{*kk}^{(c_*)}=(\lambda_0^{(c_*)})^{2k+1},\quad
A_{*k,k+2}^{(c_*)}=-2i\alpha_2\frac{\sqrt{(k+1)(k+2)}}{W}\,\Big(1+O\Big(\frac{k+1}{W}\Big)\Big),\\
&|A_{*k,k+2p}^{(c_*)}|\le \frac{C^p(k+1)^p}{W^p}.
\label{pA.4}\end{align}
In addition, if  $\{\widetilde\psi_k\}$ are defined by (\ref{pA.1}) with $c_*$ replaced by some $c_0>0$,  
$\tilde P_l$ is a projection on  the space, spanned on $\{\tilde\psi_r\}_{k=0}^l$, and $P_m$ is a similar projection for $\{\psi_k\}_{k=0}^m$,
then for  any $l,m>2$
\begin{align}\label{P_lm}
\|\tilde P_l(1-P_m)\|\le Cl^3/m
\end{align}
with $C$, depending only on $c_*$ and $c_0$.
\end{lemma}
The proof of Lemma \ref{l:A_*} is given in Appendix.

Choose $W,n$-independent $\delta>0$, which is small enough to provide that the domain
$\omega_\delta=\{x\in\mathbb{R}: |F(x)|>1-\delta\}$ contains two non intersecting sub domains  $\omega_\delta^{+}$, $\omega_\delta^{-}$, 
such that each of $\omega_\delta^{+}$, $\omega_\delta^{-}$ contains one of the points $x=a_+$ and $x=a_-$ of maximum $F(x)$
(easier speaking,  $\omega_\delta^{+}$, $\omega_\delta^{-}$ are two non-intersecting neighbourhood of points $a_+$ and $a_-$).

Consider the basis $\{\psi_{k,\delta}^+\}$, obtained by the
Gramm-Schmidt orthonormalization procedure of 
\[
\psi_k^+(x)=\psi_k(x-a_+)
\]
on $\omega_\delta^{+}$. Here we take $\{\psi_{k}\}_{k=0}^\infty$ of (\ref{pA.1}) with $c_*=c_+$ of (\ref{c_pm}). 
Since $\psi^+_{k,\delta}(x)=O(e^{-cW})$ for $x\not\in\omega^+_\delta$, one can obtain easily
that
\begin{align}
&\psi^+_{k,\delta}(x)=\psi_{k}^+(x)+O(e^{-cW}),\quad k\ll W.
\notag\end{align}
By the same way we construct $\{\psi_{k}^-(x)\}_{k=0}^\infty$ and $\{\psi_{k,\delta}^-(x)\}_{k=0}^\infty$ on $\omega_\delta^{-}$ (with $c_*=c_-$). Take some sufficiently large but $W$, $n$-independent $m$ and denote  $P_+$ and $P_-$ the projections on the subspaces spanned on the systems
$\{\psi_{k,\delta}^+\}_{k=0}^m$ and $\{\psi_{k,\delta}^-\}_{k=0}^m$ respectively.
 Evidently these projection operators are orthogonal to each other. Set
\begin{align}\label{L_i}
P=P_++P_-,\quad\mathcal{L}_1 =P\,L_2[\mathbb{R}], \quad\mathcal{L}_2=(1-P)\,L_2[\mathbb{R}],\quad L_2[\mathbb{R}]=\mathcal{L}_1\oplus\mathcal{L} _2.
\end{align}
In order to
apply Proposition \ref{p:sp} to $A$, we consider the operator $A$ as a block operator with respect to the decomposition (\ref{L_i}). It has the form
\begin{align}\label{tA.4}
&A^{(11)}=A_+^{(m)}+A_-^{(m)}+O(e^{-cW}),\quad A_+^{(m)}:=P_+AP_+,\quad A_-^{(m)}:=P_-AP_-;\\ \notag
&A^{(12)}=P_+A(I_+-P_+)+P_-A(I_--P_-)+O(e^{-cW});\\ \notag
&A^{(21)}=(I_+-P_+)AP_++(I_--P_-)AP_-+O(e^{-cW});\\ \notag
&A^{(22)}=(1-P)A(1-P),
\notag\end{align}
where $I_+$ and $I_-$ are the operators of multiplication by $\mathrm{1}_{\omega_\delta^{+}}$ and $\mathrm{1}_{\omega_\delta^{-}}$.
Indeed, since 
\begin{equation*}
A\psi_{k,\delta}^+(x)=\int A(x,y)\psi_{k,\delta}^+(y)dy=\int \mathcal{F}(x)e^{-W^2(x-y)^2/2}\mathcal{F}(y) \psi_{k,\delta}^+(y)dy,
\end{equation*}
$ |\mathcal{F}(x)|\le 1$, and evidently for $k\le m$  $\psi_{k,\delta}^+(y)$ is $O(e^{-cW})$ for $|y-a_+|\ge \varepsilon$ with any small $W$-independent $\varepsilon$, we get
\begin{equation}\label{razm1}
A\psi_{k,\delta}^+(x)= O(e^{-cW}),  \quad |x-a_+|\ge 2\varepsilon. 
\end{equation}
Therefore, for instance, $(P_-AP_+)f=O(e^{-cW})$, $(I_--P_-)AP_+=O(e^{-cW})$, etc., which gives (\ref{tA.4}).

Now let $\widehat{A}$ be the matrix $A$ without the row and the column corresponding to $\psi_{0,\delta}^+$, and $\widehat{A}^{(11)}$, 
$\widehat{A}^{(12)}$, and $\widehat{A}^{(21)}$ be the blocks of this matrix similar to (\ref{tA.4}). Denote also
\begin{align*}
 a=&A\psi_{0,\delta}^+-(A\psi_{0,\delta}^+,\psi_{0,\delta}^+)\psi_{0,\delta}^+\\
 a^{*}=&A^{*}\psi_{0,\delta}^+-(A^*\psi_{0,\delta}^+,\psi_{0,\delta}^+)\psi_{0,\delta}^+
\notag\end{align*}
and set
\begin{align*}
F(z)=(A\psi_{0,\delta}^+,\psi_{0,\delta}^+)-z-((\widehat{A}-z)^{-1}a,a^*)
\end{align*}
Then, according to Proposition \ref{p:sp}, to prove Theorem \ref{t:A}, it suffices to show the bounds
\begin{align}\label{b_a}
&\|a\|\le CW^{-3/2},\quad \|a^{*}\|\le CW^{-1}\\
&\|(\widehat{A}-z)^{-1}\|\le CW,\quad (A\psi_{0,\delta}^+,\psi_{0,\delta}^+)=\lambda_{0,+}+O(W^{-3/2})
\notag\end{align}
for $z$, satisfying the conditions  
\begin{align}\label{cond_z}
1-\dfrac{3\alpha_1}{2W}\le|z|\le 1,\quad |z-\lambda_{0,-}|>C/W.
\end{align}
Indeed, consider $z\in \omega_+=\{z: |z-\lambda_{0,+}|\le C_0W^{-3/2}\}$ with sufficiently big $C_0$, and set
$$
F_0(z)=\lambda_{0,+}-z.
$$
Then (\ref{b_a}) implies
\begin{multline}\label{in_Rouch}
|F(z)-F_0(z)|=|(A\psi_{0,\delta}^+,\psi_{0,\delta}^+)-\lambda_{0,+}-((\widehat{A}-z)^{-1}a,a^{*})|\\
\le O(W^{-3/2}) 
<C_0W^{-3/2}=|F_0(z)|, \quad z\in \partial \omega_+
\end{multline}
for sufficiently big $C_0$. Since both functions are analytic in $\omega_+$,  the Rouchet theorem gives that $F$  and $F_0$  have the same numbers of roots
(i.e., one) in $\omega_+$. This yields the first assertion of Theorem \ref{t:A} (for $\lambda_{0,-}$ the proof is the same). 
To prove the second assertion, consider any point $z_0$ outside of
$\omega_+$ satisfying (\ref{cond_z})  and take $\omega_0=\{z: |z-z_0|\le C_0W^{-3/2}\}$. Then (\ref{in_Rouch}) is still true on $\partial \omega_0$, and hence the number of roots of $F$ in
$\omega_0$ is the same as for $F_0$,  i.e. zero.  Therefore, $A$ has only one eigenvalue in the domain (\ref{cond_z}). Applying similar argument for $\lambda_{0,-}$
instead of $\lambda_{0,+}$, we obtain Theorem \ref{t:A} with $c_1=\alpha_1/2$.

Hence, we are left to prove (\ref{b_a}). The bound  $\|(\widehat{A}-z)^{-1}\|\le CW$ follows from 
 three lemmas.

\begin{lemma}\label{l:calG}
Given $z$ satisfying (\ref{cond_z}), we have
\begin{align*}
||(\widehat{A}^{(11)}-z)^{-1}||\le CW
\end{align*}
\end{lemma}
\begin{lemma}\label{l:A_12}
We have
\begin{align*}
&\|{A}^{(12)}\|\le CW^{-1},\;\|{A}^{(21)}\|\le CW^{-3/2},\;\\ 
&\|a^*\|\le CW^{-1},\;\| a\|\le CW^{-3/2}.\notag
\end{align*}
\end{lemma}
\begin{lemma}\label{l:A_22}
Given $z$ satisfying (\ref{cond_z}), we have
\begin{align*}
\|A^{(22)}\|\le 1-Cm^{1/3}/W,\quad\Rightarrow\quad
\|(A^{(22)}-z)^{-1}\|\le CW.
\end{align*}
\end{lemma}

\subsubsection{Proof of Lemmas \ref{l:calG} -- \ref{l:A_22}} 
\textit{Proof of Lemma \ref{l:calG}.}
According to (\ref{tA.4}), we have to prove that 
\begin{equation}\label{in_A.11}
\|(\widehat{A}_+^{(m)}-z)^{-1}\|\le CW, \quad \|(A_-^{(m)}-z)^{-1}\|\le CW,
\end{equation} 
where $\widehat{A}_+^{(m)}$ is $A_+^{(m)}$ without the line  and the column, corresponding to $\psi_{0,\delta}^+$.

Let us prove the first inequality of (\ref{in_A.11}).

Using that $\psi_{k,\delta}^+(y)$ is $O(e^{-c\log^2W})$ for $|y-a_+|\ge W^{-1/2}\log W$
 for $k\le m$ , we get similarly to (\ref{razm1})
\begin{equation}\label{razm}
A\psi_{k,\delta}^+(x)= O(e^{-c\log^2W}),  \quad |x-a_+|\ge 2W^{-1/2}\log W. 
\end{equation}
In addition,
$A\psi_{k,\delta}^+(x)$ can be written in the form ($k=0,1,\ldots,m$)
\[A\psi_{k,\delta}^+(x)=\int _{|y-a_+|\le W^{-1/2}\log W}(A_*^+(x-a_+,y-a_+)+\widetilde A(x,y))\psi_{k,\delta}^+(y)dy+O(e^{-c\log^2W}). 
\]
Here and below we denote
\begin{equation}\label{A_*^pm}
A_*^\pm:=A_*^{(c_\pm)},
\end{equation}
and  
\[\widetilde A(x,y)=A(x,y)-A_*^+(x-a_+,y-a_+).
\]
Expanding $\mathcal{F}$ for $|x-a_+|\le 2W^{-1/2}\log W$, $|y-a_+|\le 2W^{-1/2}\log W$, we get
\begin{align}\label{G.2a}
\widetilde A(x,y)=A(x,y)O(W^{-3/2}\log^3W ),\quad\mathrm{if}\,\, |x-a_+|+|y-a_+|\le W^{-1/2}\log W.
\end{align}
 Thus, for $k\le m$
\begin{align}\label{G.3}
 \widehat A_+^{(m)}=\widehat{A}^{(m)}_{+,*}
+\widetilde{A}^{(11)},\,\,\hbox{where}\,\, \widehat{A}^{(m)}_{+,*}=\{A^+_{*jk}\}_{j,k=1}^{m},\,\hbox{and}\,\,||\widetilde{A}^{(11)}||\le CW^{-3/2}.
\end{align}
Hence it suffices to prove that  for $z$ satisfying (\ref{cond_z}) we have
\begin{align}\label{G.4}
\|(\widehat{A}^{(m)}_{+,*}-z)^{-1}\|\le CW.
\end{align}
Decompose
\begin{align}\label{G.5}
\widehat{A}^{(m)}_{+,*}-z=D(z)+R,\quad \hbox{where}\quad D_{jk}=\delta_{jk}(A^+_{*kk}-z),
\end{align}
By using (\ref{pA.3}) we get that $\|D(z)^{-1}\|\le C W$ and for $j-k\not=2$ all non-zero entries $(RD^{-1})_{kj}=O(W^{-1})$ or less. In addition,  according to (\ref{cond_z})
\begin{align}\label{b_D.1}
&|D_{kk}|=|A_{*kk}^+-z|=|1-\dfrac{(2k+1)\alpha_+}{W}+O(W^{-2})-z|\\ \notag
&\ge \Big||z|-|1-\dfrac{(2k+1)\alpha_+}{W}+O(W^{-2})|\Big|\ge \dfrac{(2k-1/2-\varepsilon)\alpha_1}{W}
\end{align}
with some small fixed $\varepsilon>0$. Thus, by (\ref{pA.4})
\begin{align}\label{C<1}
|(RD^{-1})_{kk+2}|\le \frac{|\alpha_2|\sqrt{(k+1)(k+2)}}{\alpha_1(2k+7/2-\varepsilon)}+O(kW^{-1})<C<1,
 \end{align}
since $\alpha_2<\alpha_1$ in view of (\ref{alpha}) and the fact that $\arg c_\pm\in (-\pi/2,\pi/2)$.
Besides, 
$$
(RD^{-1})^m=0,
$$ 
and we get (\ref{G.4}) from
\[(\widehat{A}^{(m)}_{+,*}-z)^{-1}=\sum\limits_{l=0}^m D^{-1}(z)(RD^{-1}(z))^l.\]
Thus, we obtain the first bound of (\ref{in_A.11}) in view of (\ref{G.3}) and (\ref{C<1}). The second bound of (\ref{in_A.11}) can be obtained by the same way.
The only difference is that,  if we defined $R_-$ and $D_-$ similarly to 
(\ref{G.5}), then to prove $\|D_-(z)\|\le CW$ and $|(R_-(D_-(z))^{-1})_{02}|\le C<1$ we have to use also that $|z-\lambda_{0,-}|\ge C/W$.
$\square$
\begin{remark}\label{r:a}
Applying the Taylor expansions  up to the $m$-th order to the functions $\mathcal{F}(x)$ and $\mathcal{F}(y)$ 
 one can prove
that 
\begin{align}\label{r_a.1}
|(A_\pm^{(m)})_{jk}|\le (Cm/W)^{|j-k|/2}.
\end{align}
Indeed,  it is well known that 
the Hermite polynomials $\{\psi_k(x)\}_{k=0}^\infty$ satisfy the recursion relation
\[
 x\psi_k(x)=\sqrt{\frac{k+1}{4\alpha_1W}}\psi_{k+1}(x)+\sqrt{\frac{k}{4\alpha_1W}}\psi_{k-1}(x).
 \]
Hence,  the operator $\widehat L$ of multiplication by $x-a_+$ has a three diagonal form in the  basis $\{\psi_k^+\}$,  and 
$\widehat L^l$ has $2l+1$ non empty diagonals. The recursion relations combined with (\ref{pA.4}) yield (\ref{r_a.1}).

Bound (\ref{r_a.1}) implies, in particular, that if $a_0$ and $a^*_0$ are the parts of $a$ and $a^*$ which belong to $A^{(11)}$, then
\begin{align}\label{r_a}
\|a-a_0\|\le (Cm/W)^{m/2}, \quad \|a^*-a_0^*\|\le (Cm/W)^{m/2}.
\end{align}
\end{remark}

\textit{Proof of Lemma \ref{l:A_12}.} According (\ref{tA.4}), to prove Lemma \ref{l:A_12} we have to prove
\begin{align}\label{L3.3.1}
&\|P_+A(I_+-P_+)\|\le CW^{-1}, \quad \|P_-A(I_--P_-)\|\le CW^{-1},\\ \label{L3.3.2}
&\|(I_+-P_+)AP_+\|\le CW^{-3/2}, \quad \|(I_--P_-)AP_-\|\le CW^{-3/2}.
\end{align}
Let us prove the first inequality in (\ref{L3.3.1}) (the second is similar).
Use the bound valid for any $(m+1)\times\infty$ matrix:
\[\|M\|^2\le\sum_{j=0}^m\|M_{j}\|^2,\]
where $M_{j}=M^{*}\psi^+_{j,\delta}$.
The Parseval identity implies
\begin{align*}\|(P_+A^*(I_+-P_+))\psi_{j,\delta}^+\|^2=\|A^{*}\psi_{j,\delta}^+\|^2-\|\big(A_+^{(m)}\big)^{*}\psi_{j,\delta}^+\|^2.
\end{align*}
Using the  argument of Lemma \ref{l:calG}, we get for $j\le m$
\begin{align*}
&\|A^{*}\psi_{j,\delta}^+\|^2=\|(A_*^{+})^*\psi_{j}^+\|^2+O((m/W)^{3/2});\\
&\|\big(A_+^{(m)}\big)^{*}\psi_{j,\delta}^+\|^2=\|(A^{{(m)}}_{+,*})^*\psi_{j}^+\|^2+O((m/W)^{3/2}).
\end{align*}
Hence,  the  Parseval identity  and the bounds (\ref{pA.3}) -- (\ref{pA.4}) yield 
\begin{align*}
\|P_+A(I_+-P_+)\|\le \Big(\sum_{j=0}^m\sum_{k>m}| A^+_{*jk}|^2\Big)^{1/2}\le Cm/W+O(m(m/W)^{3/2}),
\end{align*}
which gives (\ref{L3.3.1}) (recall that $m$ is $W$-independent). The bounds (\ref{L3.3.2}) can be obtained similarly, if we use the fact that $ A^+_{*jk}=0$ for $j>k$.
$\square$

\textit{Proof of Lemma \ref{l:A_22}.}
Consider any $\|u\|=1$ such that $(u,\psi_{k,\delta}^+)=(u,\psi_{k,\delta}^-)=0$ for $k=0,1,\ldots, m$.
Split $x\in \mathbb{R}$ into  three sub domains, according to the value of the function $\mathcal{F}(x)$:
\begin{align}\label{omega}
&\Lambda_1=\{x:|\mathcal{F}(x)|\ge 1- \delta/2\};\\
&\Lambda_2=\{x:1- \delta\le |\mathcal{F}(x)|<1-\delta/2 \};\notag\\
&\Lambda_3=\{x:|\mathcal{F}(x)|<1-\delta\},\notag
\end{align}
and let $u_s$ be the projections of $u$ corresponding to $\Lambda_s$, $s=1,2,3$.

By (\ref{A})
\begin{align}\notag
||Au||^2\le ||\mathcal{F}^2u||^2\le& ||u_1||^2+(1-\delta/2)^2||u_2+u_3||^2\\
=&1-(1-(1-\delta/2)^2)||u_2+u_3||^2\notag\\
\Rightarrow&||u_2+u_3||^2\le C_0(1-||Au||^2).
\label{u_2}\end{align}
Set
\begin{align}\label{u^pm}
u_0:=u_1+u_2,\quad  u_0^{+}=u_0 1_{\omega_\delta^{+}},\quad
u_0^{-}=u_0 1_{\omega_\delta^{-}},
\end{align}  
Note that according to the choice of $u$
\begin{equation}\label{cond.LA_22a}
(u_0^{+},\psi_{k,\delta}^+)=O(e^{-cW}),\quad(u_0^{-},\psi_{k,\delta}^-)=O(e^{-cW}), \quad k=0,\dots m.
\end{equation}
\begin{lemma}\label{l:A_22a} 
For any $u_0^+$, $u_0^-$   satisfying  (\ref{cond.LA_22a}) we have
\begin{align}\label{A22.1}
||A^{(22)}u_0^{\pm}||^2\le (1-\frac{Cm^{1/3}}{W})\|u_0^{\pm}\|^2.
\end{align}
\end{lemma}
The proof of Lemma \ref{l:A_22a} is given after the end of the proof of Lemma \ref{l:A_22}. Now we assume for the moment that Lemma \ref{l:A_22a} 
is proved and finish the proof of Lemma \ref{l:A_22}.

According to (\ref{cond.LA_22a}) the assumptions of Lemma \ref{l:A_22a} are satisfied. Moreover, 
\begin{align}\label{u_3}
\Re(A u_0,Au_{3})&=\Re(A u_1,Au_{3})+\Re(A u_2,Au_{3})\\
&=O(e^{-cW^2})+\Re(Au_2, Au_{3})\le O(e^{-cW^2})
+\frac{1}{2}\|u_2+u_3\|^2,
\notag
\end{align}
since $\|A\|\le 1$. Thus we have by (\ref{u_2}),  (\ref{A22.1})
\begin{align}
||Au||^2=&\|A(u_0^{+}+u_0^{-}+u_3)\|^2\notag
\\=&||Au_0^+||^2+||Au_0^-||^2+2\Re(A u_0,Au_{3})+||Au_{3}||^2+O(e^{-cW})\notag\\
\le &(1-Cm^{1/3}/W)(\|u_0^{+}\|^2+\|u_0^{-}\|^2)+
2||u_2+u_3||^2+O(e^{-cW})\notag\\
\le &1-Cm^{1/3}/W+2C_0(1-||Au||^2)\notag\\
\Rightarrow&(1-||Au||^2)(1+2C_0)\ge Cm^{1/3}/W\notag\\
\Rightarrow &||Au||^2\le 1-C_1m^{1/3}/W.
\label{u_4}\end{align}
Here in the third line we used that
\[
\|Au_3\|^2\le \|u_3\|^2\le \|u_2+u_3\|^2.
\]
Now since, by definition, the block  ${\mathcal{A}}^{(22)}$ corresponds to $u$, which are orthogonal to $\{\psi_{k,\delta}^+\}_{k=0}^m$ and
$\{\psi_{k,\delta}^-\}_{k=0}^m$, the last inequality proves that 
\begin{align}\notag
&||A(1-P)||\le 1-C_1m^{1/3}/W\\
&\Rightarrow \|A^{(22)}\|=||(1-P)A(1-P)||\le1-C_1m^{1/3}/W\notag\\
&\Rightarrow \|(A^{(22)}-z)^{-1}\|\le |z^{-1}|\sum_{s=0}^{\infty}(\|A^{(22)}\|/|z|)^s\le C_1W/m^{1/3},
\label{u_5}\end{align}
which gives the assertion of Lemma \ref{l:A_22}. We are left to prove Lemma \ref{l:A_22a}.

$\square$

\textit{Proof of Lemma \ref{l:A_22a}.}
We prove first the relation for $u_0^{+}$. Choose $c_0>0$ sufficiently small to provide
\[\Re f(x)\ge \dfrac{c_0}{2}(x-a_+)^2,\quad x\ge 0,\quad \mathcal{F}_0=e^{-c_0(x-a_+)^2/2},\quad A_0=\mathcal{F}_0B\mathcal{F}_0\]
 Consider the basis $\{\widetilde\psi_k\}_{k\ge 0}$ in which $A_0^*A_0$ is  diagonal. The straightforward calculus gives (cf. Lemma \ref{l:A_*})
\begin{align}\label{A22.2}
&\widetilde\psi_k(x-a_+)=\widetilde h_k^{-1/2} e^{\widetilde\alpha W (x-a_+)^2}\Big(\frac{d}{dx}\Big)^k
e^{-2\widetilde\alpha W (x-a_+)^2}\\
&\widetilde \alpha=\sqrt{\dfrac{c_0}{2}}\Big(1+\frac{c_0}{2W^2}\Big)^{1/2},\quad\widetilde h_k=k!(4\widetilde\alpha W)^{k-1/2}\sqrt{2\pi},
\notag\end{align}

Now, by the assumptions of the lemma $(u_0^{+},\psi_k^+)=(u_0^{+},\psi_{k,\delta}^+)+O(e^{-cW})=O(e^{-cW})$ for $k=0,1,\ldots, m$, thus 
\[u_0^{+}=(1-P_m)u_0^{+}+O(e^{-cW}),\]
Hence,  setting $l=[m^{1/3}/\tilde C\sqrt[3]{2}]$ with $\tilde C$, depending only on $C$ in
(\ref{P_lm}), and denoting by $P_l$  the orthogonal  projection
 on the linear span of $\{\widetilde\psi_k\}_{k=0}^l$, we get by (\ref{P_lm})
\begin{align}\label{b_u_0+}
&||P_lu_0^{+}||^2\le\frac{||u_0^{+}||^2}{2}
\end{align}
Moreover,  the commutators $[\mathcal{F},B]$ and $[\mathcal{F}_0,B]$ admit the bounds
\begin{align*}
\|\,[\mathcal{F},B]\,\|\le&\sup_{x}W\int|\mathcal{F}(x)-\mathcal{F}(y)|e^{-W^2(x-y)^2}dy\\
\le & C W\int|x-y|e^{-W^2(x-y)^2}dy\le C_*/W,
\end{align*}
and similarly
\[\|[\mathcal{F}_0,B]\,\|\le C_*/W.\]
Thus
\begin{align*}
||Au_0^{+}||^2=&(\mathcal{F}^*B\mathcal{F}^*\mathcal{F}B\mathcal{F} u_0^{+},u_0^{+})\le(B\mathcal{F}^*\mathcal{F}^*\mathcal{F}\mathcal{F}B u_0^{+},u_0^{+})+2C_*W^{-1}\|u_0^{+}\|^2\\
\le&(BF_0^4B u_0^{+},u_0^{+})+2C_*W^{-1}\|u_0^{+}\|^2\le(\mathcal{F}_0B\mathcal{F}_0^2B \mathcal{F}_0u_0^{+},u_0^{+})+4C_*W^{-1}\|u_0^{+}\|^2
\\=&(A_0u_0^{+},A_0u_0^{+})+4C_*W^{-1}\|u_0^{+}\|^2=\|A_0u_0^+\|^2+4C_*W^{-1}\|u_0^{+}\|^2.
\end{align*}
Then, since $[A_0,P_l]=0$, $||A_0(1-P_l)||\le 1-l\sqrt{2c_0}/2W$ (see (\ref{pA.4})), and $\|P_lu_0^+\|\le\frac{1}{2}\|u_0^{+}\|^2$, we have
\begin{align*}
||Au_0^{+}||^2=&||A_+(P_lu_0^{+}+(1-P_l)u_0^{+})||^2+O(e^{-cW})\\
\le&||A_0(P_lu_0^{+}+(1-P_l)u_0^{+})||^2+4C_*W^{-1}\|u_0^{+}\|^2\\
=&||A_0P_lu_0^{+}||^2+||A_0(1-P_l)u_0^{+}||^2+4C_*W^{-1}\|u_0^{+}\|^2\\ 
\le&||P_lu_0^{+}||^2+(1-l\sqrt{2c_0}/2W)(\|u_0^{+}\|^2-||P_lu_0^{+}||^2)+4C_*W^{-1}\|u_0^{+}\|^2\\
\le&(1-l\sqrt{2c_0}/2W)\|u_0^{+}\|^2+||P_lu_0^{+}||^2l\sqrt{2c_0}/2W+4C_*W^{-1}\|u_0^{+}\|^2\\
\le &\big(1-l\sqrt{2c_0}/4W+2C_*W^{-1}\big)\|u_0^{+}\|^2\le (1-l\sqrt{2c_0}/5W)\|u_0^{+}\|^2
\end{align*}
if $l$ is sufficiently large.
$\square$

\subsection{Analysis of $K_*$}
 \begin{proposition}\label{p:K(U)}
If we  consider
 $K_*(t,U_1,U_2)$ of (\ref{K(U)}  as a kernel of the self-adjoint integral operator in $L_2[\mathring{U}(2),\mu(U)]$,
  its eigenvectors  $\{\phi_{\bar j}(U)\}$ ($\bar j=(j,k)$, $j=0,1,\ldots$, $k=-j,\ldots, j$) do not depend on $a_1,a_2,b_1,b_2$ and the corresponding
 eigenvalues $\{\lambda_{\bar{j}}(t)\}$ have the form
 \begin{align}\label{la_0}
\lambda_{\bar 0}(t)=1-e^{-W^2t},
 \end{align}
 and  for $t>d>0$ we have
\begin{align}\label{K.1}
&\lambda_{\bar{j}}(t)=(1-e^{-W^2t})\Big(1-\frac{j(j+1)}{W^2t}(1+O(j^2/W^2t)\Big), \\
&(\nu,\phi_0)=0.
\notag\end{align}
\end{proposition}
The proof of the proposition is given in Appendix.

It follows from Proposition \ref{p:K(U)} that, if we introduce the basis in $L_2[\mathbb{R}^2,p]\times L_2[\mathring{U}(2),dU]$
\begin{align*}
&\Psi_{\bar k,\bar{j}}(a,b,U)=\Psi_{\bar k}(a,b)\phi_{\bar{j}}(U), \\
&\Psi_{\bar k}(a,b)=\sqrt{\dfrac{2}{\pi}}\,(a-b)^{-1}\psi_{k_1}(a)\psi_{k_2}(b),
\notag\end{align*}
where $\{\psi_{k}(x)\}_{k=0}^\infty$ is some basis in $L_2[\mathbb{R}]$, then the matrix of
$K$ of (\ref{K(U)}) in this basis has a ``block diagonal  structure", which means that
\begin{align}\label{block}
&(K\Psi_{\bar k',\bar{j}},\Psi_{\bar k,\bar j_1})_p=0,\quad j\not=j_1\\
&(K\Psi_{\bar k'\bar j},\Psi_{\bar k,\bar j})_p=(K_j\Psi_{\bar k'},\Psi_{\bar k})_p\notag\\
=&\int \lambda_{\bar{j}}(t)A(a_1,a_2)A(b_1,b_2)\psi_{k_1}(a_1)\psi_{k_2}(b_1)
\psi_{k_1'}(a_2)\psi_{k_2'}(b_2)da_1db_1da_2db_2.
\notag\end{align}

\section{Analysis of $K$}

\begin{theorem}\label{t:K} For the operators $K, K(\xi)$ of (\ref{K}), (\ref{Ker})  there is an absolute $\varepsilon>0$ such that
\begin{align}\label{t2.1}
&\lambda_0(K)=\lambda_{0,+}\lambda_{0,-}+O(e^{-c\log^3W}),\quad |\lambda_1(K)|\le |\lambda_0(K)|-\varepsilon/W^2,\\
&|\lambda_0(K)-\lambda_0(K(\xi))|\le C((W/n)^2+1/n\sqrt{W}),\notag\\
&|\lambda_1(K(\xi))|\le |\lambda_{0,+}\lambda_{0,-}|-\varepsilon/2W^2,
\notag\end{align}
where $\lambda_{0,\pm}$ are defined in (\ref{tA.1}).
\end{theorem}
In particular, Theorem \ref{t:K} gives (\ref{main_1}), (\ref{main_2}), hence the assertion of Theorem \ref{thm:new}.
\subsection{Proof of Theorem \ref{t:K}}

Choose $W,n$-independent $\delta>0$, which is small enough to provide that the domain
$$\Omega_\delta=\{X: |\mathcal{F}(X)|>1-\delta\}$$ contains three non intersecting sub domains $\Omega_\delta^{\pm}$, $\Omega_\delta^{+}$, $\Omega_\delta^{-}$, 
such that each of $\Omega_\delta^{+}$, $\Omega_\delta^{-}$ contains one of the points $X_+=a_+I$ and $X_-=a_-I$ of maximum $\mathcal{F}(X)$, and
$\Omega_\delta^{\pm}$ contains the ``surface" $X^*(U)=UDU^*$ with  $D=\hbox{diag}\,\{a_+,a_-\}$, and $U\in \mathring{U}(2)$ (see (\ref{st_points_1})). 
Set also 
\[m=[\log^2 W].\]
Consider a system of functions 
\begin{align}\label{sys_pm}
&\{\Psi_{\bar k,\bar j,\delta}\}_{|\bar k|\le m,j\le (mW)^{1/2}},\\
&\bar k=(k_1,k_2),\,\,|\bar k|=\max\{k_1,k_2\},\quad \bar j=(j,l),
\,\, l=-j,\ldots, j,
\notag\end{align}
 obtained by the  Gramm-Schmidt procedure from
\begin{equation*}
\{1_{\Omega_{\delta}^{\pm}}\Psi_{\bar k,\bar j}\}_{|\bar k|\le m;j\le (mW)^{1/2}},
\end{equation*}
where
\begin{align}\label{Psi}
&\Psi_{\bar k,\bar{j}}(a,b,U)=\Psi_{\bar k}(a,b)\phi_{\bar{j}}(U), \\
&\Psi_{\bar k}(a,b)=\sqrt{\dfrac{2}{\pi}}\,(a-b)^{-1}\psi^+_{k_1}(a-a_+)\psi^-_{k_2}(b-a_-).
\notag\end{align}

Similarly, consider the system of functions $\{\Psi^{+}_{\bar k,\delta}\}_{|\bar k|\le m}$ (with $\bar k=(k_1,k_2,k_3,k_4),$ $|\bar k|=\max\{k_i\}$)
 obtained by the Gramm-Schmidt procedure from
\begin{equation*}
\{1_{\Omega_{\delta}^{+}}\, \psi^+_{k_1}(a-a_+)\psi^+_{k_2}(b-a_+) \psi^{+}_{k_3}(x)\psi^{+}_{k_4}(y)\}_{|\bar k|\le m},
\end{equation*}
 and define
$\{\Psi^{-}_{\bar k,\delta}\}_{|\bar k|\le m}$ by the same way. Denote $P_{\pm}$,  $P_+$, and $P_-$ the projections on the subspaces, spanned on these three systems.
Evidently these three projection operators are orthogonal to each other. Set
\begin{align}\label{P_i}
P=P_\pm+P_++P_-,\quad\mathcal{L}_1 =P\mathcal{H}, \quad\mathcal{L}_2=(1-P)\mathcal{H},\quad \mathcal{H}=\mathcal{L}_1\oplus\mathcal{L} _2,
\end{align}
where $\mathcal{H}$ is defined in (\ref{H}).

Besides, note that for any $f$, supported in some domain $\Omega$, and any $C>0$ (cf. (\ref{razm}))
\begin{equation}\label{razm_K}
(Kf)(X)=O(e^{-cW^2})\,\,\hbox{for}\, X: \hbox{dist}\{X,\Omega\}\ge C>0.
\end{equation}
Now consider the operator $K$ as a block operator with respect to the decomposition (\ref{P_i}).  It has the form
\begin{align}\label{K_21.0}
&K^{(11)}=K_\pm+K_++K_-+O(e^{-cW}),\\ \notag
& K_\pm:=P_{\pm}KP_{\pm},\quad K_{+}=P_{+}KP_{+},\quad K_-:=P_{-}KP_{-},\\
&K^{(12)}=P_{\pm}K(I_{\pm}-P_{\pm})+P_{+}K(I_{+}-P_{+})+P_{-}K(I_{-}-P_{-})+O(e^{-cW}), \notag\\
& K^{(21)}=(I_{\pm}-P_{\pm})KP_{\pm}+(I_{+}-P_{+})KP_{+}+(I_{-}-P_{-})KP_{-}+O(e^{-cW}),
\notag\end{align}
where $I_{\pm}$, $I_+$, and $I_-$ are  operators of  multiplication by $1_{\Omega_{\delta}^{\pm}}$,
 $1_{\Omega_{\delta}^{+}}$, and  $1_{\Omega_{\delta}^{-}}$ respectively. Indeed,
 it is easy to see from (\ref{razm_K}) and from the relation
\[
\psi_k(x)=O(e^{-cW}) \,\,\hbox{for}\, |x|\ge C>0,\quad k\le m
\]
 that, e.g. , $P_+KP_-f=O(e^{-cW})$, $P_{\pm}K(I_+-P_+)f=O(e^{-cW})$, etc (cf. (\ref{tA.4})).
 
Moreover,  by (\ref{block}) the block $K_\pm$ also  has a block diagonal structure:
 \begin{align}\label{K_11.0}
&K_\pm=\sum_{j=0}^{(mW)^{1/2}}K_{\pm}^{(j)},\quad K_{\pm}^{(j)}=\mathcal{P}_jP_{\pm}KP_{\pm}\mathcal{P}_j.
\end{align}
Here and below we denote by $\mathcal{P}_j$ the projection on linear span of $\{\Psi(a,b)\phi_{(j,l)}(U)\}_{l=-j}^j$.

Take some $W,n$-independent sufficiently small $\varepsilon$, and consider $z$, satisfying the conditions
\begin{align}\label{cond_z.1}
z\in \mathcal{D}_\varepsilon:=\{z:|\lambda_{0,+}\lambda_{0,-}|-\varepsilon W^{-2}\le |z|\le 1 \wedge |z-\lambda_{0,+}\lambda_{0,-}|>\varepsilon W^{-2} \}
\end{align}
Following Proposition \ref{p:sp}, introduce also the vectors 
\begin{align}\label{kappa}
\kappa=K\Psi_{\bar0,0,\delta}-(K\Psi_{\bar0,0,\delta},\Psi_{\bar0,0,\delta})\Psi_{\bar0,0,\delta},\quad
\kappa^*=K^\dagger\Psi_{\bar0,0,\delta}-(K^\dagger\Psi_{\bar0,0,\delta},\Psi_{\bar0,0,\delta})\Psi_{\bar0,0,\delta}.
\end{align}
The proof of Theorem \ref{t:K} is based on the following three lemmas.
 \begin{lemma}\label{l:K_11}  
 Given $z\in \mathcal{D}_\varepsilon$ consider $G^{(j)}(z)=(K_{\pm}^{(j)}-z)^{-1} $.
 Then each $G^{(j)}$ can be represented in the form (\ref{p_sp.3}), and the corresponding matrices $\widehat G_j$ and functions $F_j$ of (\ref{sp.1})
 satisfy the bound
\begin{align}\label{K_11.1}
&\|\widehat G^{(j)}\|\le CmW,\quad |\widehat G^{(j)}_{\bar k\bar k'}|\le CmWq^{|\bar k-\bar k'|/2},\quad 0<q<1,\\
&|F_j(z)|>C W^{-2},\quad || G^{(j)}||\le CW^2.
\notag\end{align}
In addition,
\begin{align}\label{K_11.1a}
&\|( K_+-z)^{-1}\|\le CW,\quad \| (K_--z)^{-1}\|\le CW.
\end{align}
\end{lemma}
\begin{lemma}\label{l:K_21} Fot the off-diagonal blocks of the operator $K$ (see (\ref{K_21.0})) we have
\begin{align}\label{K_21.1}
&||K^{(21)}||\le Cm^{3/2}/W^{3/2},\quad ||K^{(12)}||\le Cm/W,\\
&\|\kappa\|\le CW^{-3/2},\quad \|\kappa^*\|\le CW^{-1}.
\notag\end{align}

Moreover, there is some absolute $p>0$ such that
\begin{align}\label{GK_12}
\|(\widehat K^{(11)}-z)^{-1}\widehat K^{(12)}\|\le Cm^p.
\end{align}
\end{lemma}
\begin{lemma}\label{l:K_22} 
\begin{align}\label{K_22.0}
||K^{(22)}||\le 1-Cm^{1/3}/W
\end{align}
\end{lemma}
Defer the proofs of the lemmas to the next section and
 show  how one can finish the proof of Theorem \ref{t:K}, using the lemmas.

It is easy to see that the first line of (\ref{K_21.1}) and (\ref{K_22.0}) yield
\begin{align}\label{GK_21}
\|( K^{(22)}-z)^{-1} K^{(21)}\|\le Cm^pW^{-1/2}.
\end{align}
Here and below we denote by $p$ some absolute exponents which could be different in different formulas.
The  bound and (\ref{GK_12}) imply that
$\widehat K-z$ can be represented in the form
\[
\widehat K-z=\left(\begin{array}{cc}\widehat K^{(11)}-z &0\\0& K^{(22)}-z\end{array}\right)\left(\begin{array}{cc} I &O(m^p)\\ O(m^pW^{-1/2})& I
\end{array}\right).
\]
Both matrices here are invertible, and the inverse of the second one has a similar form. Hence
\[
\widehat G(z)=(\widehat K-z)^{-1}=\left(\begin{array}{cc} I\cdot (1+o(1)) &O(m^p)\\ O(m^pW^{-1/2})& I\cdot (1+o(1))\end{array}\right)
\left(\begin{array}{cc} (\widehat K^{(11)}-z)^{-1} &0\\0& (K^{(22)}-z)^{-1}\end{array}\right).
\]
Thus we get from the representation and (\ref{K_11.1}) 
\begin{align*}
\|\widehat G\|\le CW^{2}.
\end{align*}
Moreover, if we set $\widetilde G=(\widehat K+\widetilde{K}^{0}-z)^{-1}$, where
 $\widetilde{K}^{0}$ is $\widetilde K$ without  the ``line" and the ``column", corresponding to $\Psi_{\bar 0,0,\delta}$, then
 taking into account that $\|\widetilde{K}^{0}\|\le C/n\ll W^{-2}$, we obtain 
\begin{align}\label{h_0}
\widetilde G=\widehat G (1+\widetilde{K}^{0}\widehat G)^{-1}=\widehat G (1+H_0\widehat G),\quad \|H_0\|\le n^{-1}, 
\end{align} 
 and so for $W^2\ll n$
 \begin{align}\label{lG.2}
 \|\widetilde G\|\le CW^{2}.
\end{align} 
Notice that  the definition of $\Psi_{\bar 0,0,\delta}$ (see (\ref{sys_pm}), (\ref{Psi})), and $\mathcal{P}_j$ (see (\ref{K_11.0})) 
and  (\ref{kappa}), combined with (\ref{r_a}) yield
\begin{align}\notag
&\qquad\kappa=\mathcal{P}_0\kappa,\quad \kappa^*=\mathcal{P}_0\kappa^*,\\
&\Rightarrow  \hat G\kappa=\hat G^{(0)}\kappa+O((CW)^{-m}),\quad
\hat G\kappa^*=\hat G^{(0)}\kappa^*+O(e^{-c\log^3W}).
\label{Gk}\end{align}
Consider the function $F_K(z)$ of the form (\ref{sp.1}), constructed for $K$. Then  (\ref{block}) and (\ref{la_0}) yield
\begin{align}\label{F_K}
F_K(z):=&(K\Psi_{\bar0,0,\delta},\Psi_{\bar0,0,\delta})-((\widehat K-z)^{-1}\kappa,\kappa^*)\\
=&(A_{+}^{(m)})_{00}(A_{-}^{(m)})_{00}-z-(\widehat G^{(0)} \kappa_0,\kappa_0^*)+O(e^{-c\log^3W})\notag\\
=&:F_{\pm}(z)+O(e^{-c\log^3W}),
\notag\end{align}
where $A^{(m)}_+$ and $A^{(m)}_-$ are defined in (\ref{tA.4}).
But since $F_{\pm}$ is constructed according to Proposition \ref{p:sp} for $A_{+}^{(m)}\otimes A_{-}^{(m)}$, we know
that $F_{\pm}(z)$ has no roots in $D_\varepsilon$ and has exactly one root in $ \omega_\varepsilon=\{z: |z-\lambda_{0,+}\lambda_{0,-}|\le \varepsilon W^{-2}\}$. In addition 
\begin{align}\label{dF_pm}
&\frac{d}{dz}F_{\pm}(z)=-1+((\widehat G^{(0)})^2 \kappa_0,\kappa_0^*)=-1+O(W^{-1/2})\\
&\Rightarrow |F_\pm(z)|>\varepsilon W^{-2}/2,\,\, z\in\mathcal{D}_\varepsilon
\notag\end{align}
Hence, by the Rouchet theorem,  we conclude that $F_K(z)$ has no roots in $\mathcal{D}_\varepsilon$ and has exactly one root in $\omega_\varepsilon$.
This gives us the first line of (\ref{t2.1}).

To apply Proposition \ref{p:sp} to $K+\widetilde K$, denote  by $\tilde\kappa$ and $\tilde\kappa^*$   the vectors
\[
\tilde\kappa=\widetilde K\Psi_{\bar0,0,\delta}-(\widetilde K\Psi_{\bar0,0,\delta},\Psi_{\bar0,0,\delta})\Psi_{\bar0,0,\delta},\quad
\tilde\kappa^*=\widetilde K^*\Psi_{\bar0,0,\delta}-(\widetilde K^*\Psi_{\bar0,0,\delta},\Psi_{\bar0,0,\delta})\Psi_{\bar0,0,\delta}
\]
Then by the second line of (\ref{K.1})
\begin{align}\label{ti_kappa}
\mathcal{P}_0\tilde\kappa=0,\quad\mathcal{P}_0\tilde\kappa^*=0.
\end{align}
Moreover, we have
\begin{align}\label{kappa1.2}
&||\tilde\kappa||=O(n^{-1}),\quad || \tilde\kappa^*||=O(n^{-1}).
\end{align}
By Proposition \ref{p:sp}, one should study zeros of the function
\begin{align*}
\widetilde F_K(z)=&((K+\widetilde K)\Psi_{\bar 0,0},\Psi_{\bar 0,0})-z-
(\widetilde  G(z)(\kappa+\tilde\kappa),(\kappa^*+\tilde\kappa^*)).
\end{align*}
Let us prove that there is $C>0$ such that
\begin{align}\label{de_F}
|F_K(z)-\widetilde F_K(z)|\le C\big( (W/n)^2+1/n\sqrt{W}+O(e^{-c\log^3W})\big).
\end{align}
By  (\ref{lG.2}) and (\ref{kappa1.2}) we have, 
\begin{align}\notag
&|(\widetilde G\tilde\kappa,\tilde\kappa^*)|\le \|\widetilde G\|/n^2\le CW^2/n^2.
\end{align}
Moreover, the representation (\ref{h_0}) and relations (\ref{Gk}), (\ref{ti_kappa}), and  (\ref{K_11.1})  yield
\begin{align}
|(\widetilde G\kappa,\tilde\kappa^*)|&\le|(\widehat G\kappa,\tilde\kappa^*)|+
|(\widehat GH_{0}\widehat G\kappa,\tilde\kappa^*))|\label{Gk,ti_k}\\
&\le |(\widehat G^{(0)}\kappa,\tilde\kappa^*)|+|(\widehat GH_{0}\widehat G^{(0)} \kappa,\tilde\kappa^*)|+O(e^{-c\log^3W}/n)\notag\\
&\le  ||\widehat G||\cdot||H_{0}||\cdot||\widehat G^{(0)}||\cdot||\kappa||\cdot||\tilde\kappa^*||+O(e^{-c\log^3W}/n)\notag \\
&\le C_2W^{3/2}/n^2+O(e^{-c\log^3W}/n).
\notag\end{align}
Here we have used that 
\[
\widehat G^{(0)}\kappa=\mathcal{P}_0\widehat G^{(0)}\kappa\Rightarrow (\widehat G^{(0)}\kappa,\tilde\kappa^*)=0.
\]
Similarly to (\ref{Gk,ti_k}) we get
\begin{align*}
&|(\widetilde G\tilde\kappa,\kappa^*)|\le C_3(W/n)^2+O(e^{-c\log^3W}/n), \\
& |((\widetilde G-\widehat G)\kappa,\kappa^*)|\le C_4/n\sqrt{W}.
\end{align*}
Besides, by (\ref{K.1}),
\[(\widetilde K\Psi_{\bar0,0},\Psi_{\bar0,0})=O(n^{-2}).
\]
Thus,  (\ref{de_F}) is proved.

 Let $z_0$ be a root of $F_K(z)$.
Then (\ref{F_K}) and (\ref{dF_pm}) gives us 
for any $z: |z-z_0|=2C\big((W/n)^2+1/n\sqrt{W}\big)$:
\begin{align}
\label{Rou}|F_K(z)|\ge 2C\big((W/n)^2+1/n\sqrt{W}\big)(1-O(W^{-1/2}))>|\widetilde F_K(z)-F_K(z)|
\end{align}
Then, by the Rouchet theorem, $\widetilde F_K(z)$ has exactly one root in the circle 
\[|z-z_0|\le 2C\big((W/n)^2+1/n\sqrt{W})\big)\]
and has no roots in $\mathcal{D}_\varepsilon$, defined in (\ref{cond_z.1}).

$\square$

\subsection{Proofs of Lemmas \ref{l:K_11}--\ref{l:K_22}}

\textit{Proof of Lemma \ref{l:K_11}.} Let us first prove (\ref{K_11.1}). To this end we use representation of $K$ in polar coordinates (see (\ref{rep_2}) -- (\ref{K(U)})).
According to (\ref{K.1})
\begin{align*}\notag
&\lambda_{\bar j}(t)=\lambda^*_j+O(j^2(t-(a_+-a_-)^2)/W^2),
\end{align*}
where $t$ is defined in (\ref{t}), and
\begin{align}
&\lambda^*_j=1-\frac{j(j+1)}{W^2(a_+-a_-)^2}.
\label{lam*}\end{align}

\begin{definition} \label{d:O_*} We will denote by $O_*((m/W)^{3/2})$ any operator $ T$ satisfying the following property:  there exist $p_1,p_2>0$ such that
\begin{align}\label{O_*}
||TP_{\pm}||\le Cm^{p_1}W^{-3/2}, \quad |(TP_{\pm})_{\bar k,\bar k'}|\le m^{p_2} (Cm/W)^{|\bar k-\bar k'|}.
\end{align}
\end{definition}
Using (\ref{block}) and the fact that, according to Remark \ref{r:a}, the operator $\tilde T$ of multiplication by 
$(t-(a_+-a_-)^2)$   is $O_*((m/W)^{3/2})$, we obtain that
\[  K^{(j)}_{\bar k,\bar k'}=\lambda_j^*\cdot (K^{(0)}_{\bar k,\bar k'}+O_*((m/W)^{3/2})=\lambda_j^*\cdot (A^{(m)}_{+*}\otimes A^{(m)}_{-*}+O_*((m/W)^{3/2}).
\]
Hence
\begin{align}\label{pt.0}
\widehat G^{(j)}(z)=(\lambda_j^*)^{-1}\cdot (\widehat{A^{(m)}_{+*}\otimes A^{(m)}_{-*}}+O_*((m/W)^{3/2})-z_j)^{-1}, \quad z_j=z/\lambda^*_j,
\end{align}
where $\widehat{A^{(m)}_{+*}\otimes A^{(m)}_{-*}}$ means the matrix $A^{(m)}_{+*}\otimes A^{(m)}_{-*}$ without the line and the column, corresponding
to $\psi_{0}^+\otimes\psi_{0}^-$.
\begin{proposition} \label{p:tens} Take  $\mathcal{A}=A^{(m)}_{\sigma_1*}\otimes A^{(m)}_{\sigma_2*}\otimes\dots \otimes A^{(m)}_{\sigma_s*}$ with
$\sigma_i=+,-$, $i=1,\dots, s$ and $s$ independent of $n,W,m$.  Let
\begin{align}\label{pt.z}
|z-\lambda_0(\mathcal{A})|\le c_0/W^2\vee |z|>|\lambda_0(\mathcal{A})|+C_0/W,
\end{align}
$\widehat{\mathcal{A}}$ be ${\mathcal{A}}$ without the line and the column, corresponding
to $\psi_{0}^{\sigma_1}\otimes\dots\otimes\psi_{0}^{\sigma_s}$, and $\,\widehat{\mathcal{G}}(z)=(\widehat{\mathcal{A}}-z)^{-1}$,
$F(z)$ be defined as in (\ref{sp.1}). 

Then there is $0<q<1$ such that
\begin{align}\label{pt.1}
&\|\widehat{\mathcal{G}}(z)\|\le CmW,\quad |(\widehat{\mathcal{G}}(z))_{\bar k,\bar k'}|\le Wm^sq^{|\bar k-\bar k'|/2},\\
&|F(z)|\ge C|z-\lambda_0(\mathcal{A})|^{-1}.
\notag\end{align}
\end{proposition} 
The proof of the proposition is given in Appendix.

Using (\ref{pt.0}) and the proposition, we obtain
\[\widehat G^{(0)}(z)=\widehat{\mathcal{G}}\cdot (1+O_*((m/W)^{3/2}))^{-1}=\widehat{\mathcal{G}}+O_*((m/W)^{3/2})
\]
This and (\ref{p_sp.3}) gives (\ref{K_11.1}) (for $j>0$ the proof is similar).

To prove (\ref{K_11.1a}) for $K_+$, we use the representation of $K(\xi)$ in the form (\ref{rep_1}) -- (\ref{A_1}) (for $K_-$ the proof is similar). 
Note that we have for $|\bar k|\le m$ (cf. (\ref{razm}))
\begin{equation*}
(K\Psi_{\bar k,\delta}^+) (X)=O\big(e^{-C\log^2 W}\big), \quad \hbox{dist}\{X, X_+\}\ge 2W^{-1/2}\log W,
\end{equation*}
and so (similarly to the proof of Lemma \ref{l:calG}) we can write for $\hbox{dist}\{X, X_+\}\le 2W^{-1/2}\log W$, $\hbox{dist}\{X, X_+\}\le 2W^{-1/2}\log W$
\begin{align*}
(P_+A_1P_+)(X,Y)=A_{+,*}^{(m)}(x_1,x_2)A_{+,*}^{(m)}(y_1,y_2)\big(1+ O_*((m/W)^{3/2})\big).
\end{align*}
Thus, to prove the first bound of (\ref{K_11.1a}), it suffices to prove that 
\[
\|(A_{+,*}^{(m)}\otimes A_{+,*}^{(m)}\otimes A_{+,*}^{(m)}\otimes A_{+,*}^{(m)}-z)^{-1}\|\le CW,
\]
which follows from (\ref{pt.1}) in the case, when the second condition of (\ref{pt.z}) is valid, and (\ref{p_sp.3}).

$\square$

\textit{Proof of Lemma \ref{l:K_21}.}
Using (\ref{block}) and (\ref{lam*}), by the same argumet as in Lemma \ref{l:K_11}, we get
\begin{align}\label{K_12.*}
&(P_{\pm}K(I_{\pm}-P_{\pm}))_{(\bar k,\bar j),(\bar k',\bar j')}\\
&=\delta_{\bar j \bar j'}\lambda_j^*\cdot \Big((P_{+m}\otimes P_{-m})(A^+\otimes A^-)(1-P_{+m}\otimes P_{-m})\Big)_{\bar k,\bar k'}+
(O_*((m/W)^{3/2}))_{\bar k,\bar k'}
\notag\\
&=\Big(\sum_{j\le (mW)^{1/2}}\Big(\lambda_j^*\cdot (P_{+m}\otimes P_{-m})\big(A^+\otimes A^-)(1-P_{+m}\otimes P_{-m}\big)
\notag\\&\hskip8cm +O_*((m/W)^{3/2})\Big)\otimes\mathcal{P}_j
\Big)_{(\bar k,\bar j),(\bar k',\bar j')},
\notag\end{align}
where $P_{+m}$ and $P_{-m}$ are defined by the same way as $P_m$ in Lemma \ref{l:A_*} for the operators $A_*^+$ and $A_*^-$ respectively.
  Hence, using that 
  \begin{align*}
 1 -P_{+m}\otimes P_{-m}=P_{+m}\otimes(1- P_{-m})+(1-P_{+m})\otimes P_{-m}
-(1-P_{+m})\otimes(1- P_{-m}).
  \end{align*}  
  we obtain on the basis of Lemma \ref{l:A_12} :        
\begin{align*}
&\|P_{\pm}K(I_{\pm}-P_{\pm})\|\le \|(P_{+m}\otimes P_{-m})(A^+\otimes A^-)\big(1-P_{+m}\otimes P_{-m}\big)\|+O(m^p/W^{3/2})\\
\le&\| \big(P_{+m}A^+P_{+m}\big)\otimes\big(P_{-m}A^-(1-P_{-m})\big)\|+\|\big(P_{+m}A^+(1-P_{+m})\big)\otimes \big(P_{-m}A^-P_{-m}\big)\|
\\&\qquad \qquad+\|\big(P_{+m}A^+(1-P_{+m})\big)\otimes \big(P_{-m}A^-(1-P_{-m})\big)\|+O(m^p/W^{3/2})\\
\le&C(\|P_{+m}A^+(1-P_{+m})\|+\|P_{-m}A^-(1-P_{-m})\|)+O(m^p/W^{3/2})\le CmW^{-1}.
\end{align*}

By the same way one can prove the bound for  $(I_{\pm}-P_{\pm})KP_{\pm}$, $P_+K(I_+-P_+)$, $P_-K(I_- -P_-)$,
$(I_+-P_+)KP_+$, and $(I_- -P_-)KP_-$.
The  second  line of (\ref{K_21.1})  evidently follows from the first one combined with (\ref{r_a.1}). 

To prove (\ref{GK_12}),  denote by $R_j$ the $j$th operator in the r.h.s.  of (\ref{K_12.*}) (including the error term)
and each  $R_j$  split  into two parts:
\begin{align}\label{split_R}
&R_j=R_{0j}+R_{1j},\quad \|R_{1j}\|\le C(m/W)^2, \quad \|R_{0j}\|\le C(m/W),\\
& (R_{0j}\Psi_{\bar k',j},\Psi_{\bar k,j})=0,\;\mathrm{if}\; \;\,|\bar k'|>m+3.
\notag\end{align}
For this aim we set
$$R_{1j}=\lambda_j^*\Big((P_{+m}\otimes P_{-m})\big(A^+\otimes A^-)(1-P_{+(m+3)}\otimes P_{-(m+3)}\big)\otimes\mathcal{P}_j\Big)
+O_*((m/W)^{3/2}),
$$
where $P_{+(m+3)}$  and $P_{-(m+3)}$ are defined as $P_{+m}$ and $P_{-m}$ with $m$ replaced by $m+3$. Then, 
using the same argument as above, we obtain  the bound (\ref{split_R}) for $||R_{1j}||$.
Setting 
$R_{0j}=R_j-R_{1j}$,
we obtain the second line of (\ref{split_R}).

Note that 
by (\ref{K_21.0}) and (\ref{K_12.*}), 
to prove (\ref{GK_12}), 
  it suffices  to check the relations
\begin{align}\label{GK_12.1}
&\|(K_+-z)^{-1}P_+K(I_+-P_+)\|\le Cm^p,\quad\|(K_--z)^{-1}P_-K(I_--P_-)\|\le Cm^p,\\
&\|\widehat G^{(0)}\widehat R_0\|\le Cm^p,\quad
\| G^{(j)}R_j\|\le Cm^p,\quad j>1,
\label{GK_12.2}\end{align}
with $R_j$,  defined above, and $ \widehat R_0$, being $R_0$ without the line, corresponding to $\Psi_{\bar 0,0,\delta}$.

The bounds of (\ref{GK_12.1})  follow from  (\ref{K_11.1a}) and the second line of  (\ref{K_21.1}).
The first bound  of (\ref{GK_12.2}) follows from the first bound of (\ref{K_11.1}) and
(\ref{K_12.*}). To obtain the second one, we  use   (\ref{split_R}).
The bounds of the norms of $R_{1j}$ and $G^{(j)}$ yield
\[
|| G^{(j)}R_{1j}||\le Cm^p,\quad j>0. 
\]
Hence, to prove (\ref{GK_12}), we are left only to check that
$|| G^{(j)}R_{0j}||\le Cm^p$.

By (\ref{p_sp.3}) and the bound for $F_j(z)$ from (\ref{K_11.1})
it suffices to prove that  if we denote by $R_{0j}^{(\bar k')}$  the column of $R_{0j}$ with a number $\bar k'$, then
\begin{align}\label{GK_12.3}
&|(\widehat G^{(j)}\kappa^*_j,R_{0j}^{(\bar k')})|\le Cm^p/W^2,\\
&\hbox{for} \,\,\bar k'=(m+\alpha,k'_2)\vee \bar k'=(k'_1,m+\alpha),\,\, (\alpha=1,2,3,\,\,k_1',k_2'\le m).
\notag\end{align}
Consider the case $k'=(m+1,k'_2)$ (other ones are similar). Using that $|\bar l|\ge (|l_1|+|l_2|)/2$, we get
\begin{align*}
&|( R_{0j}^{(\bar k')})_{\bar k}|
 =|\lambda_j^* \cdot (A_{+}^{(m)})_{k_1,m+1'} (A_{-}^{(m)})_{k_2k_2'}|+O((Cm/W)^{(m+1-k_1)/4})\\
 &\le (Cm/W)|^{((m+1-k_1)+|k_2-k_2'|)/4}.
\end{align*}
By the definition (\ref{sp.1}) and the bounds (\ref{r_a.1}), we have
\begin{align*}
|(\kappa^*_{\bar j})_{\bar k}|
=| \lambda_j^* \cdot (A_{+})_{0k_1} (A_{-})_{0k_2}|+O((Cm/W)^{(k_1+k_2)/4})\le(Cm/W)^{(k_1+k_2)/4}.
\end{align*}
Now, using (\ref{K_11.1}), it is easy to obtain that
\[|(\widehat G^{(j)}\kappa^*_j,R_{0j}^{(\bar k')})|\le CW^2m^pq^{m/4}\ll W^{-2}\]
To estimate $|| G^{(j)}R_{1j}||$, one can just sum the bounds for different $k_2'\le m$ and add similar bounds for the other
cases of (\ref{GK_12.3}).
Thus, we  proved (\ref{GK_12.3}) and hence finished the proof of  (\ref{GK_12}).

$\square$

\textit{Proof of Lemma \ref{l:K_22}.}
Let us split  the integration domain $X\in \mathcal{H}$ into  3 sub domains, according to the value of the function $\mathcal{F}(X)$ (cf. (\ref{omega})):
\begin{align}\label{Omega}
&\Lambda_1=\{X:|\mathcal{F}(X)|\ge 1- \delta/2\},\\
&\Lambda_2=\{X:1-\delta/2>|\mathcal{F}(X)|\ge 1- \delta\},\notag\\
&\Lambda_3=\{X:1-\delta>|\mathcal{F}(X)|\},\notag
\end{align}
Then, similarly to the proof of Lemma \ref{l:A_22}, write
\[u(X)=u_1(X)+u_2(X)+u_3(X),
\]
where $u_i(X)=u(X)1_{X\in\Lambda_i}$. Since $\max_{X\in\Lambda_2\cup\Lambda_3}|\mathcal{F}(X)|=1-\delta$,
we have similarly to (\ref{u_2}) for any $u:\|u\|=1$
\begin{align}\label{ineq1}
||u_2+u_3||^2\le C_0(1-||{K}u||^2).
\end{align}
Moreover, similarly to (\ref{u_3})
\begin{align}\label{ineq2}
\Re(K(u_1+u_2),Ku_3)\le O(e^{-cW^2})+\frac{1}{2}||u_2+u_3||^2.
\end{align}  
Hence, if we denote 
\begin{align}\label{u^pm_0}
u_0:=u_1+u_2,\quad u_0^{(\pm)}=u_0 1_{\Omega_\delta^{(\pm)}},\quad u_0^{(+)}=u_0 1_{\Omega_\delta^{(+)}},\quad
u_0^{(-)}=u_0 1_{\Omega_\delta^{(-)}},
\end{align}  
and  prove 
for $u_0^{(\pm)}$,  $u_0^{(+)}$, and $u_0^{(-)}$ the analogue of Lemma \ref{l:A_22a}, then repeating 
the bounds of (\ref{u_4})-(\ref{u_5}), we obtain the bound
for ${K}^{(22)}$.
\begin{lemma}\label{l:K_22a}
For $u_0^{(\pm)}$,  $u_0^{(+)}$, and $u_0^{(-)}$ defined in (\ref{u^pm_0}) we have
\begin{align}\label{K_22a.1}
&\|Ku_0^{(\pm)}\|^2\le 1-Cm^{1/3}/W, \\ \notag
&\|Ku_0^{(-)}\|^2\le 1-Cm^{1/3}/W,\quad \|Ku_0^{(+)}\|^2\le 1-Cm^{1/3}/W.
\end{align}
\end{lemma}
{\textit Proof.}  To prove the first bound, let us note first that, by the assumption of the lemma,
\[
u_0^{(\pm)}=u_{01}^{(\pm)}+u_{02}^{(\pm)}+O(e^{-cW}),
\]
where
\begin{align*}
&u_{01}^{(\pm)}=\sum_{j\le (mW)^{1/2}}\sum_{|\bar k|> m}u_{\bar k,j}\Psi_{\bar k,j},\quad
u_{02}^{(\pm)}=\sum_{j> (mW)^{1/2}}\sum_{\bar k}u_{\bar k,j}\Psi_{\bar k,j}.
\end{align*}
Hence,
\[
(Ku_{01}^{(\pm)},Ku_{02}^{(\pm)})_p=0\Rightarrow \|Ku_{01}^{(\pm)}+Ku_{02}^{(\pm)}\|^2= \|Ku_{01}^{(\pm)}\|^2+\|Ku_{02}^{(\pm)}\|^2.
\]
Moreover, it is easy to see that the space $\mathcal{L}$ spanned on the functions 
$\{\Psi_{\bar k}(a,b)\phi_j(U)\}_{\bar j>(mW)^{1/2}}$
is invariant with respect to $K$ and for $u\in \mathcal{L}$
\[
\|Ku\|^2\le \max_{j>(m W)^{1/2}}(\lambda_ju,\lambda_j u)_p\le (1-Cm^2/W)\|u\|^2
\]
Thus
\begin{align}\label{K_22a.2}
\|Ku_{02}^{(\pm)}\|^2\le(1-Cm^2/W)\|u_{02}^{(\pm)}\|^2.
\end{align}
To obtain a similar bound for $\|Ku_{01}^{(\pm)}\|^2$, we use the same  method, as in the proof of Lemma \ref{l:A_22a}.
Consider the operator kernel $A_{0,+}(a_1,a_2):=A_0(a_1,a_2)$, defined as in the proof of Lemma \ref{l:A_22a}, and 
similarly define $A_{0,-}(a_1,a_2)$ (with $a_-$ instead of $a_+$).
Set
\[
K_0(a_1,a_2,b_1,b_2,U_1,U_2)=t^{-1}A_{0,+}(a_1,a_2)A_{0,-}(b_1,b_2)K_*(t,U_1,U_2)
\]
Choose $\widetilde\Psi_{\bar k}(a,b)=(a-b)^{-1}\widetilde\psi_{k_1}(a-a_+)\widetilde\psi_{k_2}(b-a_-)$, $\bar k=(k_1,k_2)$. Then  evidently
\[
(\widetilde \Psi_{\bar k},\Psi_{\bar k'})_p=(\widetilde\psi_{k_1},\psi_{k_1'})(\widetilde\psi_{k_2},\psi_{k_2'}).
\]
By our choice of $u_{01}^{(\pm)}$
\[
u_{01}^{(\pm)}=\sum_{|\bar k|\ge m}u_{\bar k}(U)\Psi_{\bar k}+O(e^{-cW}).
\]
Hence,   setting $l=[m^{1/3}]/3 C^{1/3}$ (with ${C}$ of (\ref{P_lm})) and denote by $\widetilde P_l$ the orthogonal projection operator on the linear  span of $\{\widetilde\Psi_{\bar k}\}_{|\bar k|\le l}$,
  we get 
\begin{align}\notag
&\|\widetilde P_lu_{01}^{(\pm)}\|^2\le\int dU\sum_{|\bar k|\le l}|(u_{01}^{(\pm)},\widetilde\Psi_{\bar k})_p |^2\\
&=\int dU\sum_{|\bar k|\le l}\Big|\Big(\sum_{k'_1\ge m,k_2'} +\sum_{k'_2\ge m,k_1'}-\sum_{k'_1,k_2'\ge m}\Big)
u_{\bar k'}(U)(\psi_{k_1'},\tilde\psi_{k_1})(\psi_{k_2'},\tilde\psi_{k_2})\Big|^2.
\label{*}\end{align}
Let us show how to estimate the first sum with respect to $\bar k'$ above. 
Denote
\[v_{k_1'}=\sum_{k_2'}u_{\bar k'}\psi_{k_2'}.
\]
Then, by the Schwartz inequality and (\ref{b_k,j}) (see below), we obtain
\begin{align*}
\Sigma_1=&\int dU\sum_{k_1,k_2\le l}\Big|\sum_{k_1'>m}(v_{k_1'},\tilde\psi_{k_2})(\psi_{k_1'},\tilde\psi_{k_1})\Big|^2\\
&\le\int dU
\Big(\sum_{k_2\le l}\sum_{k_1'>m}|(v_{k_1'},\tilde\psi_{k_2})|^2\Big)
\Big(\sum_{k_1\le l}\sum_{k_1'>m}|(\psi_{k_1'},\tilde\psi_{k_1})|^2\Big)\\
&\le \tilde C \frac{l^3}{m}\sum_{k_1'}\int dU (v_{k_1'},v_{k_1'})\le \tilde C \frac{l^3}{m}\|u^{(\pm)}\|^2
\le \|u^{(\pm)}\|^2/27.
\end{align*}
Using similar bounds for the second and the third sum of (\ref{*}) and denoting the respective expression by
$\Sigma_2$ and $\Sigma_3$, we get
\[\|\widetilde P_lu_{01}^{(\pm)}\|^2\le 3\Sigma_1+3\Sigma_2+3\Sigma_3\le  \|u^{(\pm)}\|^2/3<\|u^{(\pm)}\|^2/2.
\]
Then, repeating the argument of Lemma \ref{l:A_22a}, we get that
\[
\|Ku_{01}^{(\pm)}\|^2\le(1-Cm^{1/3}/W)\|u_{01}^{(\pm)}\|^2
\]
and  finish the proof of the lemma for $u_0^{(\pm)}$.
For $u_0^{(+)}$ and $u_0^{(-)}$ the proofs are similar.

$\square$

\section{Appendix}
\textbf{Proof of Lemma \ref{l:A_*}.}
Orthonormality of $\{\psi_k\}_{k\ge 0}$ follows from the orthonormality of $\{p_k\}_{k\ge 0}$;
(\ref{pA.0}) can be easily checked by the straightforward calculations.

Let us compute
\begin{align*}
A_*^{(c_*)}\psi_k&=\dfrac{h_k^{-1/2}W}{\sqrt{2\pi}}\int e^{-W^2(x-y)^2/2-c_*(x^2+y^2)/2}\cdot e^{-\alpha W y^2}e^{2\alpha_1Wy^2}\Big(\dfrac{d}{dy}\Big)^ke^{-2\alpha_1Wy^2} d y\\
&=\dfrac{h_k^{-1/2}W}{\sqrt{2\pi}}e^{W^2x^2/2d-(W^2+c_*^2)x^2/2}\int e^{-W^2d(y-x/d)^2/2}\Big(\dfrac{d}{dy}\Big)^ke^{-2\alpha_1Wy^2} dy,
\end{align*}
where 
$$
d=1+c_*/W^2-2\bar\alpha/ W.
$$ 
Integration by parts gives
\begin{align*}
A_*^{(c_*)}\psi_k
&=\dfrac{(-1)^kh_k^{-1/2}W}{\sqrt{2\pi}}e^{W^2x^2/2d-(W^2+c_*^2)x^2/2}\int \Big(\dfrac{d}{dy}\Big)^ke^{-W^2d(y-x/d)^2/2}e^{-2\alpha_1Wy^2} dy\\
&=\dfrac{d^kh_k^{-1/2}W}{\sqrt{2\pi}}e^{W^2x^2/2d-(W^2+c_*^2)x^2/2}\Big(\dfrac{d}{dx}\Big)^k\int e^{-W^2d(y-x/d)^2/2-2\alpha_1Wy^2} dy,
\end{align*}
and then we obtain by the straightforward  calculations  
\begin{align}\label{pA.2}
&A_*^{(c_*)}\psi_k=h_k^{-1/2}\cdot\lambda_0^{(c_*)}\cdot\Big(1-\frac{2\bar\alpha}{W}+\frac{c_*}{W^2}\Big)^k\cdot
e^{d_2 Wx^2}\Big(\frac{d}{dx}\Big)^ke^{-d_1W x^2},\\
&d_1=\frac{2\alpha_1}{(1-\frac{2\bar\alpha}{W}+\frac{c_*}{W^2})(1+\frac{2\alpha}{W}+\frac{c_*}{W^2})}
=2\alpha_1\Big(1-\frac{4i\alpha_2}{W}\Big)+O(W^{-2}),\notag\\
& d_2=d_1-\alpha. 
\notag\end{align}
According to (\ref{pA.2}), we have
\begin{align*}
&A_*^{(c_*)}\psi_k=\widetilde p_k(x)e^{-\alpha Wx^2},\quad \widetilde p_k(x)=\gamma_k^{(k)}x^k+\gamma_{k-2}^{(k)}x^{k-2}+\ldots,\\
 &\gamma_k^{(k)}=\lambda_0^{(c_*)}\cdot  h_k^{-1/2} (-2d_1W)^{k}d^k, \quad \gamma_{k-2}^{(k)}=\dfrac{k(k-1)}{2}\cdot \lambda_0^{(c_*)} \cdot h_k^{-1/2} (-2d_1W)^{k-1}d^k.
\end{align*}
Moreover,  we can obtain from orthonormality of $\psi_k$
\begin{align*}
&\int x^l p_k (x) e^{-2\alpha_1Wx^2}dx=0,\quad l<k,\\
&\int x^k p_k (x) e^{-2\alpha_1Wx^2}dx=h_k^{1/2} (-4\alpha_1 W)^{-k},\\
&\int x^{k+2} p_k (x) e^{-2\alpha_1Wx^2}dx=-h_k^{1/2} (-4\alpha_1 W)^{-k-1}\cdot \dfrac{(k+2)(k+1)}{2}.
\end{align*}
Hence,
\begin{align*}
(A_*^{(c_*)}\psi_k,\psi_k)&=\int \widetilde p_k(x) p_k(x) e^{-2\alpha_1W x^2}=
\int \gamma_k^{(k)}x^k \cdot p_k(x)e^{-2\alpha_1 Wx^2}dx\\
&= d^k (d_1/2\alpha_1)^{k}\lambda_0^{(c_*)}=(\lambda_0^{(c_*)})^{2k+1}.
\end{align*}
Similarly,  we can write
\begin{align*}
(A_*^{(c_*)}\psi_{k+2},\psi_{k})&=
\int \widetilde p_{k+2}(x) p_{k}(x) e^{-2\alpha_1W x^2}d x\\ 
&=\int (\gamma_{k+2}^{(k+2)}x^{k+2}+\gamma_k^{(k+2)} x^k)p_{k}(x) e^{-2\alpha_1W x^2}d x\\
&=\dfrac{\sqrt{(k+2)(k+1)}}{2}\Big(\dfrac{d_1}{2\alpha_1}\Big)^{k+1}\Big(\dfrac{d_1}{2\alpha_1}-1\Big) \cdot \lambda_0^{(c_*)} \cdot d^{k+2}, 
\end{align*}
which gives (\ref{pA.3}). By the same argument one can obtain (\ref{pA.4}),
using that
\[
\dfrac{\sqrt{(k+1)(k+2)\ldots (k+2l)}}{l!}\le C^l(k+1)^l.
\]
To prove (\ref{P_lm}), let us prove first that for  $j,k>2$
\begin{align}\label{b_k,j}
|(\psi_j,\widetilde\psi_k)|\le C(k/j).
\end{align}
Indeed, changing $x\to x/\sqrt{W}$, we get
\begin{align*}
(\psi_j,\widetilde\psi_k)=(\varphi_j,e\widetilde\varphi_k),
\end{align*}
where
\begin{align}\label{def_phi}
\varphi_j(x)=&h_{j0}^{-1/2}\Big(\dfrac{d}{d x}\Big)^je^{-2\alpha_1x^2},\quad h_{j0}=j! (4\alpha_1)^{j-1/2} \sqrt{2\pi},\\
\widetilde\varphi_k(x)=&\widetilde h_{k0}^{-1/2}\Big(\dfrac{d}{d x}\Big)^ke^{-2\widetilde\alpha x^2},\quad
\widetilde h_{k0}=k! (4\widetilde\alpha)^{k-1/2} \sqrt{2\pi},\notag\\
e(x)=&e^{(\overline{\alpha}+\widetilde\alpha) x^2}.
\notag\end{align}
Since
\[\varphi_j(x)=(4\alpha_1)^{-1}(j(j-1))^{-1/2}\varphi_{j-2}''(x),\]
and similar relations are valid for $\widetilde\varphi_k(x)$, integration by parts yelds
\begin{align}\notag
(\varphi_j,e\widetilde\varphi_k)&= (4\alpha_1)^{-1}(j(j-1))^{-1/2}(\varphi_{j-2},(e\varphi_k)'')\\ 
&=(4\alpha_1)^{-1}(j(j-1))^{-1/2}\Big(\varphi_{j-2},e''\widetilde\varphi_k+2e'(\alpha(k+1))^{1/2}\varphi_{k+1}\notag\\
&\hspace{5cm}+e\cdot 4\widetilde\alpha((k+2)(k+1))^{1/2}4\widetilde\alpha\varphi_{k+2}\Big).
\label{int_parts}
\end{align}
Since $\varphi_ke^{{\alpha_1}x^2}$ and $\widetilde\varphi_je^{\widetilde\alpha x^2}$ by definition (\ref{def_phi}) are the normalized Hermite functions,
the above relation proves (\ref{b_k,j}).
Now,   let $u=(1-P_m)u$, hence
\[u=\sum_{k> m}u_k\psi_k.
\]
Then,  by (\ref{b_k,j}),
\begin{align*}
&||P_lu||^2=\sum_{k=0}^l|(u,\widetilde\psi_k)|^2= \sum_{k=0}^l\Big|\sum_{j> m}u_j(\psi_j,\widetilde\psi_k)\Big|^2
\le||u||^2
 \sum_{k=0}^l\sum_{j> m}|(\psi_j,\widetilde\psi_k)|^2\\ \notag
 &\le C||u||^2\sum_{k\le l}\sum_{j> m}\frac{k^2}{j^2}\le \dfrac{C||u||^2}{m}\sum_{k\le l}k^2\le C||u||^2\frac{l^3}{m}
\end{align*}

$\square$

\textit{Proof of Proposition \ref{p:tens}}.
Similarly to the proof of Lemma \ref{l:calG}, consider
the diagonal matrix with the entries
\[D_{\bar k\bar k}=A^{\sigma_1}_{*k_1k_1}\dots A^{\sigma_s}_{*k_sk_s}-z. 
\]
As in (\ref{b_D.1}), we have
\[
|D_{\bar k\bar k}|>2\alpha_1(k_1+\dots +k_s+s/2-\varepsilon)/W.
\]
Set also $R=(\widehat K_{\pm}^{(0)}-D-z)D^{-1}$ and let $Q$ be the matrix which contains  $O(1)$-order entries of
$R$ while the other ones are replaced by zeros. This gives
\begin{equation}\label{r_tild}
(\widehat K_{\pm}^{(0)}-z)^{-1}=D^{-1}(I+R)^{-1}=D^{-1} (1+Q)^{-1}(I+\widetilde R)^{-1},
\end{equation}
where $\widetilde R=(R-Q)(I+Q)^{-1}$.

It follows from (\ref{r_a.1})  that, if we set $e_1=(1,0,\dots,0),\dots e_s=(0,\dots,0,1)$, then
\begin{align}\label{Q} 
Q_{\bar k \bar k'}\not=0,\,\mathrm{iff}\, \bar k'=\bar k+2e_1\vee\dots\vee \bar k'=\bar k+2e_s 
\end{align}
Moreover, there exists an absolute constant $l_{\alpha}$, such that for $|k|>l_\alpha$
\begin{align*} 
&\sum_{i=1}^s|Q_{\bar k,\bar k+2e_i}|\le 
\sum_{i=1}^s\frac{|A^{\sigma_i}_{*k_i,k_i+2}|}{2\alpha_1(k_1+\dots+k_s+s/2)}
\\
\le&\frac{\alpha_2}{\alpha_1}\frac{\sqrt{(k_1+1)(k_1+2)}+\dots+\sqrt{(k_s+1)(k_s+2)}}{(k_1+\dots+k_s+s/2-\varepsilon)}\le (\alpha_2/\alpha_1)^{1/2}=q<1.
\end{align*}
Here we used that $\alpha_2<\alpha_1$ (see (\ref{C<1})).

Write $Q$ as a block matrix   
\begin{align*}& Q^{(11)}=\{Q_{\bar k,\bar k'}\}_{|k|\le l_{\alpha},|k'|\le l_{\alpha}},\quad
Q^{(12)}=\{Q_{\bar k,\bar k'}\}_{|k|\le l_{\alpha},|k'|> l_{\alpha}},\\
& Q^{(21)}=\{Q_{\bar k,\bar k'}\}_{|k|> l_{\alpha},|k'|\le l_{\alpha}}\quad
Q^{(22)}=\{Q_{\bar k,\bar k'}\}_{|k|> l_{\alpha},|k'|> l_{\alpha}}.
\end{align*}

 Then by (\ref{Q}) $Q^{(21)}=0$, and by (\ref{norm_A}) $||Q^{(22)}||\le q$. Moreover, (\ref{Q}) implies that for $s_0=[l_{\alpha}/2]+1$
 \[Q^{s_0}=\left(\begin{array}{cc}0&X\\0&(Q^{(22)})^{s_0}\end{array}\right)\Rightarrow
Q^{s_0+p}=\left(\begin{array}{cc}0&X(Q^{(22)})^{p}\\0&(Q^{(22)})^{s_0+p}\end{array}\right),\quad p>0,
 \]
where $X$ is some fixed matrix. 
Writing  the Neumann series $(1+Q)^{-1}=\sum_s (-1)^sQ^s$ and 
taking into account that
by  (\ref{Q})  $(Q^s)_{\bar k\bar k'}=0$, till $s<|\bar k-\bar k'|/2$, we obtain
that
 \begin{align}\label{b_Q}
& |(1+Q)^{-1}_{\bar k\bar k'}|\le Cq^{|\bar k-\bar k'|/2}
\end{align}
Besides, it follows from (\ref{r_a.1}) that $R-Q=O_*((m/W)^{3/2})$, hence
 \begin{align*}
&|(\widetilde R)_{\bar k\bar k'}|=\Big|\sum_{|\bar k''\le m}(R-Q)_{\bar k\bar k''}
(1+Q)_{\bar k'',\bar k'}\Big|\le C(m^{5/2}/W^{-1/2})q^{|\bar k-\bar k'|/2},
\end{align*}
The last relation implies
\begin{align}
 |(1+\widetilde R)^{-1}_{\bar k\bar k'}|\le Cq^{-|\bar k-\bar k'|/2}.
\label{b_R}\end{align} 
To prove this, let us consider any fixed $\bar k$ and $\bar k'$ and  use the standard trick from the spectral theory
(see e.g. \cite{PS:11}, Ch. 13.3). Assume that $|\bar k-\bar k'|=k_1-k_1'$. Then denote $ D_q$ the diagonal matrix
such that $( D_q)_{\bar k''\bar k'''}=\delta_{\bar k''\bar k'''}q^{k''_1/2}$. Then
 \begin{align*}
& ||D_q\widetilde RD_q^{-1}||\le Cm^{7/2}/W^{-1/2}\\
\Rightarrow & |(1-\widetilde R)^{-1}_{\bar k\bar k'}|=
 |(D_q(1-D_q\widetilde RD_q^{-1})^{-1}D_q^{-1})_{\bar k\bar k'}|\\
& \le q^{(k_1-k_1')/2}||(1-D_q\widetilde RD_q^{-1})^{-1}||.
\end{align*} 
If $|\bar k-\bar k'|=-(k_1-k_1')$ we use $D_q^{-1}$ instead of $D_q$. And if $|\bar k-\bar k'|=\pm(k_2-k_2')$ we use
$( D_q)_{\bar k''\bar k'''}=\delta_{\bar k''\bar k'''}q^{\pm k''_2/2}$.

Now (\ref{b_Q}), (\ref{b_R}), and (\ref{r_tild})
conclude the proof of  the first line of (\ref{pt.1}). The second line follows from the fact that 
\[F(\lambda_0(\mathcal{A}))=0 \quad  F'(z)=1-(\widehat{\mathcal{G}}^2\kappa,\kappa^*)=1-O(m^pW^{-1/2}).\]

$\square$

\textit{Proof of Proposition \ref{p:K(U)}.} 
Notice that, if for any $V\in \mathring{U}(2) $ we define an operator
\[(T_Vf)(U)=f(UV),
\]
then for any kernel of the form $\mathcal{K}(U_1,U_2)=\mathcal{K}(U_1U_2^*)  $ we have evidently
\begin{align*}
&T_V\mathcal{K}f(U)=\int\mathcal{K}(UVU_1^*)f(U_1)dU_1=\int\mathcal{K}(UU_2^*)f(U_2V)dU_2=\mathcal{K}T_Vf(U)\\
&\Rightarrow T_V\mathcal{K}=\mathcal{K}T_V
\end{align*}
Since $T_V$ is a representation of the group $\mathring{U}(2)$ in $L_2[\mathring{U}(2),dU]$, it can be represented as a sum of irreducible
representations in the subspaces $E_j$ ($L_2[dU]=\oplus E_j$). And the commutation property guarantees that $\mathcal{K}$ acts like an
 identity operator multiplied by some constant 
in each of the  subspace $E_j$. These constants $\lambda_j$ are  eigenvalues of  $\mathcal{K}$, and choosing any basis in  $E_j$ we
obtain all eigenvectors of $\mathcal{K}$.

In the standard parametrization 
\begin{equation}\label{U_param}
U=\left(
\begin{array}{cc}
\cos \varphi&\sin\varphi\cdot e^{i\theta}\\
-\sin\varphi\cdot e^{-i\theta}&\cos\varphi
\end{array}
\right),
\end{equation}
the measure $dU$ has the form
\[
dU=\dfrac{1}{\pi}u\,du\, d\theta, \quad u=|\sin\varphi|\in [0,1], \quad \theta \in [0,2\pi).
\]
Then the spaces $E_j$ of the irreducible representations are well-known, and the proper basis in $E_j$ is made from the standard spherical harmonics 
 $\phi_{\bar j}(U)$ with $\bar j=(j,k)$, $j=0,1,\ldots$, $k=-j,\ldots, j$ be 
\begin{equation}\label{phi_j}
\phi_{\bar j}(U)=l_{j,k}\, P_j^k(\cos 2\phi)\, e^{ik\theta}=l_{j,k} \Big(\dfrac{d}{dx}\Big)^k P_j(x)\Big|_{x=1-2|U_{12}|^2}(2\bar{U}_{11} U_{12})^k,
\end{equation}
where $P_j^k$ is an  associated Legendre polynomial
\begin{align}\label{Leg}
&P_j^k(\cos x)=(\sin x)^k  \Big(\dfrac{d}{d\cos x}\Big)^k P_j(\cos x),\quad P_j(x)=\dfrac{1}{2^jj!}\dfrac{d^j}{dx^j}(x^2-1)^j,\\ \notag
&l_{j,k}=\sqrt{\dfrac{(2j+1) (j-k)!}{(j+k)!}}.
\end{align}
Then $\{\phi_{\bar j}(U)\}$ is an orthonormal basis (see, e.g., \cite{Leg}, \S 12.6); 
to find $\lambda_j$ it suffices to apply our $K_*$ to $P_j$. We get
\begin{multline}\label{K*_int}
\lambda_j(t)=(K_*\phi_{(j,0)},\phi_{(j,0)})=\int K_*(t, U_1,U_2) \phi_{(j,0)}(U_2)\phi_{(j,0)}(U_1)dU_1 dU_2\\
=tW^2\int e^{-tW^2|(U_1U_2^*)_{12}|^2}f(U_2) \phi_{(j,0)}(U_2)\phi_{(j,0)}(U_1)dU_1 dU_2\\
=tW^2\int e^{-tW^2|U_{12}|^2}
\phi_{(j,0)}(U_1U^*) \phi_{(j,0)}(U_1)dUdU_1
\end{multline}
For the parametrization (\ref{U_param}) we have
\[\phi_{(j,0)}(U)=P_j(1-2|U_{12}|^2)\Rightarrow \phi_{(j,0)}(U_1U^*)=P_j(1-2|(U_1U^*)_{12}|^2). \]
Since the Legendre addition theorem (see, e.g., \cite{Leg}, \S 12.8) yields
\begin{multline}\label{Leg_ad}
P_j(1-2|(U_1U^*)_{12}|^2)=P_j(\cos 2\varphi)\cdot P_j(\cos 2\varphi_1)\\
+2\sum\limits_{l=1}^j\dfrac{(l-j)!}{(l+j)!}\,P_j^l(\cos 2\varphi)\cdot P_j^l(\cos 2\varphi_1)\cdot \cos(k(\theta-\theta_1)),
\end{multline}
integrating first with respect to $\theta_1$, we obtain that the sum above gives a zero contribution to the integral (\ref{K*_int}) and
\[\int dU_1 \phi_{(j,0)}(U)\phi_{(j,0)}(U_1U^*)=P_j(\cos 2\phi)\]
Thus,
\begin{align*}
\lambda_{\bar j}(t)=
2tW^2\int_{0}^1 e^{-tW^2u^2}P_j(1-2u^2)\, u du=\int\limits_0^{tW^2}e^{-u}\, P_j\Big(1-\dfrac{2u}{tW^2}\Big) du,
\end{align*}
which gives  the first line of (\ref{K.1}), since
\begin{align*}
\dfrac{d^l}{dx^l} P_j(x)\Big|_{x=1}&=\dfrac{1} {2^j j!} \dfrac{d^{j+l}}{dx^{j+l}}(x^2-1)^j\Big|_{x=1}=\dfrac{1}{2\pi i}\cdot \dfrac{(j+l)!}{2^jj!}\oint \dfrac{(z^2-1)^j}{(z-1)^{j+l+1}}dz\\
&=\dfrac{(j+l)(j+l-1)\ldots (j-l+1)}{2^ll!}.
\end{align*}
The second line of  (\ref{K.1}) can be obtained easily from the direct computations.

$\square$
\medskip

\textbf{Acknowledgement.}
We are  grateful to Sasha Sodin, who drew our attention to the transfer matrix approach in application to 1d random band matrices, for many
fruitful discussion. Also we would like to thank Leonid Pastur for his helpful suggestions on the proof of Proposition~\ref{p:K(U)}.
An essential part of this work was done during our stay at the Simons Center of Geometry and Physics, which we would like to thank for its hospitality.

\end{document}